\documentstyle[psfig]{mn}

\begin{document}

\journal{To Appear in MNRAS}

\title[X-ray evidence for solar abundances in the brightest groups]
{X-ray evidence for multiphase hot gas with nearly solar Fe abundances
in the brightest groups of galaxies} 
\author[D.~A. Buote]{David A. Buote$^{1,2}$\\ $^1$ UCO/Lick
Observatory, University of California at Santa Cruz, Santa Cruz, CA
95064, U.S.A. \\ $^2$ {\sl Chandra} Fellow\\ }

\maketitle

\begin{abstract}

We analyze the {\sl ASCA} spectra accumulated within $\sim 100$ kpc
radii of 12 of the brightest groups of galaxies. Upon fitting
isothermal models (1T) jointly to the {\sl ASCA} SIS and GIS spectra
we obtain fits for most groups that are of poor or at best marginal
quality and give very sub-solar metallicities similar to previous
studies, $\langle Z\rangle = 0.29\pm 0.12$~$Z_{\sun}$. Two-temperature
models (2T) provide significantly better fits for 11 out of the 12
groups and in every case have metallicities that are substantially
larger than obtained for the 1T models, $\langle Z\rangle = 0.75\pm
0.24$~$Z_{\sun}$. Although not very well constrained, for most of the
groups absorption in excess of the Galactic value is indicated for the
cooler temperature component of the 2T models. A simple multiphase
cooling flow model gives results analogous to the 2T models including
large metallicities, $\langle Z\rangle = 0.65\pm 0.17$~$Z_{\sun}$. The
nearly solar Fe abundances and also solar $\alpha$/Fe ratios indicated
by the 2T and cooling flow models are consistent with models of the
chemical enrichment of ellipticals, groups, and clusters which assume
ratios of Type Ia to Type II supernova and an IMF similar to those of
the Milky Way.

Thus, we have shown that the very sub-solar Fe abundances and Si/Fe
enhancements obtained from most previous studies within $r\sim 100$
kpc of galaxy groups are an artifact of their fitting isothermal
models to the X-ray spectra which also has been recently demonstrated
for the brightest elliptical galaxies. Owing to the importance of
these results for interpreting X-ray spectra, in an appendix we use
simulated {\sl ASCA} observations to examine in detail the ``Fe bias''
and ``Si bias'' associated with the spectral fitting of ellipticals,
groups, and clusters of galaxies.

\end{abstract}

\begin{keywords}
galaxies: general -- galaxies: evolution -- X-rays: galaxies.
\end{keywords}
 
\section{Introduction}
\label{intro}

The hot gas in elliptical galaxies, groups, and clusters of galaxies
is thought to be enriched with metals by material ejected by
supernovae. Hence, the abundances in the hot gas record the star
formation history of these systems and therefore probe the rates of
Type Ia and Type II supernovae as well as the shape of the IMF (for
reviews see Renzini 1997, 1999). Since the hot gas in these systems
emits at temperatures $T\sim 10^{7}$-$10^{8}$ $^{\circ}K$ ($\sim 1$-10
keV), X-ray observations are the preferred means to determine the
abundances in the hot gas.

Spatial analyses of the X-ray surface brightness profiles of groups of
galaxies using {\sl ROSAT} \cite{rosat} indicate that the hot gas
typically extends out to distances in excess of 100 kpc (e.g. David et
al. 1994; Pildis, Bregman, \& Evrard 1995; Davis et al. 1996; Ponman
et al. 1996; Trinchieri, Fabbiano, \& Kim 1997) and also suggest that
the gas is divided into distinct components corresponding to the halo
of the central galaxy and an extended component for the surrounding
group (Mulchaey \& Zabludoff 1998; also see Ikebe et al. 1996). These
studies typically obtain average temperatures of $T\sim 1$-2 keV from
analysis of the {\sl ROSAT} PSPC spectra and also find evidence for
positive temperature gradients in some groups (Trinchieri et al. 1997;
Mulchaey \& Zabludoff 1998). Despite the evidence for temperature
gradients, in most cases the abundances reported by the {\sl ROSAT}
studies are obtained from isothermal spectral models from which very
sub-solar Fe abundances are inferred (a notable exception being NGC
533 -- Trinchieri et al. 1997).

Similarly, analyses of the {\sl ASCA} data \cite{tanaka} of groups
which assume isothermal gas typically find very sub-solar Fe
abundances and, in some cases, Si/Fe ratios in excess of solar
(Fukazawa et al. 1996, 1998; Davis, Mulchaey, \& Mushotzky
1999). Although the point spread function (PSF) of {\sl ASCA} is quite
large and asymmetric, there have been several attempts to deconvolve
the {\sl ASCA} data to allow analysis of the spectra as a function of
radius with the results for galaxy clusters being decidedly mixed
(e.g. Markevitch et al. 1998; Kikuchi et al. 1999). Recently,
Finoguenov et al. \shortcite{fino} and Finoguenov \& Ponman
\shortcite{fp} have applied their own method of deconvolution to the
{\sl ASCA} data of a small number of groups. By assuming the hot gas
to be in a single phase, these authors have fitted isothermal models
at several radii for each group. Similar to the previous studies,
Finoguenov et al. \shortcite{fino} and Finoguenov \& Ponman
\shortcite{fp} obtain very sub-solar Fe abundances and Si/Fe ratios in
excess of solar.

The very sub-solar Fe abundances for the hot gas in groups of galaxies
deduced by these studies are in many cases less than those of galaxy
clusters and thus pose a problem for the popular models of chemical
enrichment which assume ratios of Type Ia to Type II supernova and an
IMF similar to those of the Milky Way. (e.g. Renzini 1997). To
reconcile these observations with these models it has to be postulated
that galaxy groups accrete a significant amount of primordial gas
after they have spent most of their evolution expelling gas. The
unattractive features of this scenario are discussed by Renzini
\shortcite{ren94}.

These results for groups are in every way analogous to those initially
obtained for ellipticals with {\sl ROSAT} (e.g. Davis \& White 1996)
and {\sl ASCA} (e.g. Matsumoto et al. 1997). Recently, however, Buote
\& Fabian (1998, hereafter BF) and Buote (1999, hereafter B99) have
demonstrated that models having at least two temperature components in
the hot gas are required to explain the {\sl ASCA} spectra of the
brightest ellipticals. These multiphase models agree with the positive
temperature gradients indicated by the {\sl ROSAT} data unlike the
isothermal models (B99).

Perhaps most importantly BF and B99 also showed that the Fe abundances
determined from the multiphase spectral models of the brightest
ellipticals are consistent with solar in stark contrast to the very
sub-solar values obtained from isothermal models. This strong
dependence of the Fe abundance on the spectral model indicates that
one cannot obtain a reasonable estimate of the Fe abundance by using
an isothermal model if the hot gas consists of multiple temperature
components. Similar problems also arise when determining the
$\alpha$/Fe ratios (B99).

Since the spectral properties of the brightest ellipticals are so
similar to groups, it should be expected that similar results for the
Fe abundance also apply to groups of galaxies.  To investigate this
issue we selected for analysis the groups from the samples of Mulchaey
\& Zabludoff \shortcite{mz} and Davis et al. \shortcite{dmm} having
the highest quality {\sl ASCA} data suitable for detailed spectral
analysis. In addition, we included the group NGC 1132 since Mulchaey
\& Zabludoff \shortcite{mz1132} argue that it has properties similar
to bright groups despite the fact that it has no large galaxies aside
from NGC 1132. Our final sample consists of 12 galaxies whose
properties are listed in Table \ref{tab.prop}.

The paper is organized as follows. In section \ref{obs} we present the
observations and discuss the data reduction. We define the models used
for broad-band spectral fitting\footnote{We focus on the broad-band
spectral analysis since B99 found that individual line ratios of Si
and S place only weak constraints on the spectral models of the
brightest elliptical galaxies which have data of significantly higher
S/N than the groups in our sample.} in section \ref{models} and
discuss calibration and fitting issues in section \ref{issues}. The
results for the broad-band spectral fitting analysis are presented in
section \ref{results}. We discuss the implications of our results in
section \ref{disc} and present our conclusions in section
\ref{conc}. Finally, in Appendix \ref{bias} we use simulated {\sl
ASCA} observations to examine in detail the ``Fe bias'' and ``Si
bias'' associated with the spectral fitting of ellipticals, groups,
and clusters of galaxies.

\section{{\sl ASCA} Observations and Data Reduction}
\label{obs}

\begin{table*}
\caption{Group and BGG Properties}
\label{tab.prop}
\begin{tabular}{lccccccccc}
Group & $z$ & N$_{\rm gal}$  & $\sigma$ & $N_{\rm H}$ & $\log_{10}
     L_{\rm x}$ & BGG  & Type & $B_{\rm T}^0$ & $\log_{10} L_{\rm B}$\\ 
     & & &  (km s$^{-1}$) & ($10^{21}$cm$^{-2}$) & (erg s$^{-1}$) &
     Name & & & (erg s$^{-1}$)\\   
HCG 51 & 0.0258 & $\cdots$ & 240 & 0.12 & 42.40 & NGC 3651 & E1 & 14.49 & 43.18\\
HCG 62 & 0.0146 & 45 & 376 & 0.27 & 42.20 & NGC 4761 & S0 & 13.30 & 43.16\\
MKW 9  & 0.0397 & $\cdots$ & 336 & 0.39 & 42.73 & UGC 09886 & S0 & 14.29 &  43.63\\
NGC 533  & 0.0181 & 36 & 464 & 0.32 & 42.31 & same & E3 & 12.22  & 43.78\\
NGC 1132 & 0.0232 & $\cdots$ & 253 & 0.48 & 42.48 & same & E  & 13.01 & 43.68\\
NGC 2563 & 0.0163 & 29 & 336 & 0.43 & 41.76 & same & S0 & 13.01 & 43.37\\
NGC 4104 & 0.0283 & $\cdots$ & 546 & 0.17 & 42.52 & same & S0 & 12.93 & 43.88\\
NGC 4325 & 0.0252 & 18 & 265 & 0.23 & 42.91 & same & E4 & 14.41 & 43.19\\
NGC 5129 & 0.0233 & 33 & 294 & 0.18 & 42.09 & same & E  & 12.89 & 43.73\\
NGC 5846 & 0.0061 & 20 & 368 & 0.42 & 41.90 & same & E0-1 & 11.13 & 43.47\\
NGC 6329 & 0.0276 & $\cdots$ & $\cdots$ & 0.21 & 42.07 & same & E & 13.62 & 43.59\\
RGH 80   & 0.0370 & $\cdots$ & 448 & 0.10 & 42.74 & $\cdots$ & $\cdots$ & $\cdots$ & $\cdots$\\
\end{tabular}

\medskip

\raggedright

Properties for the groups and their brightest group galaxy (BGG);
i.e. morphological types, blue magnitudes, and blue luminosities refer
to the BGG. Redshifts are taken from NED and the RC3. The number of
group galaxies (N$_{\rm gal}$) and velocity dispersions $(\sigma)$ are
taken from Zabludoff \& Mulchaey \shortcite{zm} for six
galaxies. Velocity dispersions for other galaxies are from Hickson et
al. \shortcite{hick92} for HCG 51 and 62, from Beers et
al. \shortcite{beers95} for MKW9 and NGC 4104, from Tonry \& Davis
\shortcite{td81} for NGC 1132, and from Ramella et al. \shortcite{rgh}
for RGH 80.  Galactic Hydrogen column densities $(N_{\rm H})$ are from
Dickey \& Lockman \shortcite{dl} using the HEASARC w3nh tool. The
X-ray luminosities, $L_{\rm x}$, are computed in the 0.5-10 keV energy
band from the (unabsorbed) 2T models in section \ref{2t} of this
paper. The (luminosity) distances used for these calculations are
obtained from the quoted redshifts with $H_0=70$ km s$^{-1}$ Mpc
$^{-1}$ and $\Omega_0=0.3$ except for the nearby galaxy NGC 5846 for
which use the distance from Faber et al. \shortcite{7s} rescaled for
$H_0=70$ km s$^{-1}$ Mpc $^{-1}$.  Morphological types and total
apparent blue magnitudes ($B_{\rm T}^0$) for the BGGs are from the
RC3. The blue-band luminosities of the BGGs are computed from $B_{\rm
T}^0$ using the distances as above and ${L_B}_{\sun}= 4.97\times
10^{32}$ erg s$^{-1}$.

\end{table*}

\begin{table*}
\caption{{\sl ASCA} Observations}
\label{tab.obs}
\begin{tabular}{llrcccccccccrr}
Group & Sequence & \multicolumn{1}{c}{Date} & SIS & SIS & RDD & Default & 
\multicolumn{2}{c}{SIS Temp} & SIS BKG & \multicolumn{2}{c}{Radius} & \multicolumn{2}{c}{Radius}\\
& & \multicolumn{1}{c}{Mo/yr} & CCD & Data & Correct & Screen & S0 &
S1 & & \multicolumn{2}{c}{(arcmin)} & \multicolumn{2}{c}{(kpc)}\\ 
& & & Mode & Mode & & & ($^{\circ}$C) & ($^{\circ}$C) & & SIS0 & GIS & SIS0 & GIS\\
HCG 51   & 82028000 & 6/94  & 1+2 & B & N & Y & $\cdots$ & $\cdots$ & T & 3.9 & 4.5 & 127 & 146\\
HCG 62   & 81012000 & 1/94  & 2 & B & N & Y & $\cdots$ & $\cdots$ & T & 3.0 & 3.5 & 55 & 64\\
MKW 9    & 83009000 & 2/95  & 2 & B & Y & N & -61.0 & -60.8 & L & 4.0 & 4.4 & 201 & 221\\
NGC 533  & 62009000 & 8/94  & 4 & B & N & N & -59.5 & -59.0 & L & 3.5 & 3.5 & 80 & 80\\
         & 62009010 & 1/96  & 2 & B & Y & N & -59.5 & -59.0 & L & 3.5 & 3.5 & 80 & 80\\
NGC 1132 & 65021000 & 8/97  & 1 & B & Y & Y & $\cdots$ & $\cdots$ & T & 3.8 & 4.5 & 111 & 131\\
NGC 2563 & 63008000 & 10/95 & 1 & F & Y & Y & $\cdots$ & $\cdots$ & T & 3.0 & 3.4 & 61 & 70\\
NGC 4104 & 83008000 & 12/94 & 2 & B & N & N & -60.5 & -60.0 & T & 3.0 & 4.0 & 107 & 143\\
         & 84017000 & 5/96  & 1 & F & Y & N & $\cdots$ & -61.0 & T & 3.0 & 3.4 & 107 & 121\\
NGC 4325 & 85066000 & 1/97  & 2 & F & Y & Y & $\cdots$ & $\cdots$ & T & 3.0 & 3.4 & 95 & 108\\
NGC 5129 & 84048000 & 7/96  & 2 & B & Y & Y & $\cdots$ & $\cdots$ & T & 3.0 & 3.5 & 88 & 103\\
NGC 5846 & 61012000 & 2/94  & 4 & B & N & N & $\cdots$ & $\cdots$ & L & 4.1 & 4.5 & 55 & 60\\
NGC 6329 & 84047000 & 4/96  & 2 & B & Y & Y & $\cdots$ & $\cdots$ & T & 3.0 & 3.4 & 104 & 118\\
RGH 80   & 83012000 & 6/95  & 2 & B & Y & Y & $\cdots$ & $\cdots$ & T & 3.6 & 3.8 & 169 & 178\\
\end{tabular}

\medskip

\raggedright

See text in section \ref{obs} for explanation of the terminology used
in this table. Additional information for some individual groups is
given at the end of the section. Distances computed as described in
notes to Table \ref{tab.prop}.

\end{table*}

\begin{table*}
\begin{minipage}{140mm}
\caption{{\em ASCA} Exposures and Count Rates}
\label{tab.exp}
\begin{tabular}{llrrrrrrrr}
Name  & Sequence \# & \multicolumn{2}{c}{Exposure} &
      \multicolumn{2}{c}{Count Rate} & \multicolumn{2}{c}{Exposure} &
      \multicolumn{2}{c}{Count Rate}\\
      &             &  \multicolumn{2}{c}{($10^{3}$s)}  &
      \multicolumn{2}{c}{($10^{-2}$ ct s$^{-1}$)} &
      \multicolumn{2}{c}{($10^{3}$s)}  & \multicolumn{2}{c}{($10^{-2}$
      ct s$^{-1}$)} \\
      &             & SIS0 & SIS1 & SIS0 & SIS1 & GIS2 & GIS3 & GIS2 &
      GIS3\\
HCG 51   & 82028000 & 72.0 & 71.6 & 5.7 & 4.6 & 72.0 & 72.0 & 2.2 & 2.6\\
HCG 62   & 81012000 & 17.2 & 18.7 & 10.2 & 9.5 & 33.8 & 33.8 & 3.5 & 4.7\\
MKW 9    & 83009000 & 29.0 & 18.8 & 8.1 & 5.9 & 46.7 & 46.7 & 3.9 & 4.5\\
NGC 533  & 62009000 & 8.9 & 9.4 & 5.4 & 3.4 & 19.6 & 19.6 & 2.7 & 3.6\\
         & 62009010 & 14.9 & 14.9 & 7.4 & 5.2 & 18.2 & 18.2 & 3.2 & 3.7\\
NGC 1132 & 65021000 & 30.5 & 30.5 & 8.2 & 6.1 & 29.5 & 29.4 & 2.9 & 3.6\\
NGC 2563 & 63008000 & 45.8 & 45.8 & 2.7 & 2.0 & 51.1 & 50.4 & 1.1 & 1.3\\
NGC 4104 & 83008000 & 12.2 & 11.6 & 5.8 & 4.3 & 45.8 & 45.8 & 3.4 & 4.2\\ 
         & 84017000 & 21.8 & 20.9 & 5.4 & 4.6 & 24.7 & 24.5 & 2.9 & 3.5\\
NGC 4325 & 85066000 & 19.7 & 19.7 & 18.1 & 12.7 & 29.0 & 29.0 & 5.4 & 6.1\\
NGC 5129 & 84048000 & 31.2 & 29.5 & 3.0 & 2.4 & 34.1 & 34.1 & 1.1 & 1.4\\
NGC 5846 & 61012000 & 24.8 & 11.5 & 17.7 & 9.9 & 40.4 & 40.4 & 4.8 & 5.4\\  
NGC 6329 & 84047000 & 30.9 & 32.1 & 3.2 & 2.7 & $\cdots$ & 38.3 & $\cdots$ & 1.8\\ 
RGH 80   & 83012000 & 39.2 & 37.6 & 6.0 & 4.6 & 45.2 & 44.2 & 1.9 & 2.4\\
\end{tabular}

\medskip

The count rates are given for energies 0.55-9 keV for SIS data before
1996, 0.65-9 keV after 1996, and 0.8-9 keV for the GIS. The count
rates are background subtracted within the particular aperture (see
text in section \ref{obs}).

\end{minipage}
\end{table*}

The {\sl ASCA} satellite consists of two X-ray CCD cameras (Solid
State Imaging Spectrometers -- SIS0 and SIS1) and two proportional
counters (Gas Imaging Spectrometers -- GIS2 and GIS3) each of which is
illuminated by its own X-ray telescope (XRT). The SIS has superior
energy resolution and effective area below $\sim 7$ keV while the GIS
has a larger field of view. Although the PSF of each XRT has a
relatively sharp core, the wings of the PSF are quite broad (half
power diameter $\sim 3\arcmin$) and increase markedly for energies
above a few keV (e.g. Kunieda et al. 1995). As in our previous studies
(e.g. B99) we do not attempt to analyze the spatial distribution of
the {\sl ASCA} data of these sources because of the large, asymmetric,
energy-dependent PSF. Rather, we analyze the {\sl ASCA} X-ray emission
within a single large aperture for each group which encloses the
region of the most significant emission (see below).

We obtained {\sl ASCA} data for the 12 groups from the public data
archive maintained by the High Energy Astrophysics Science Archive
Research Center (HEASARC). The properties of the observations are
listed in Tables \ref{tab.obs} and \ref{tab.exp}. We reduced these
data with the standard {\sc FTOOLS} (v4.2) software according to the
procedures described in The {\sl ASCA} Data Reduction Guide and the
WWW pages of the {\sl ASCA} Guest Observer Facility (GOF)\footnote{See
http://heasarc.gsfc.nasa.gov/docs/asca/abc/abc.html and
http://heasarc.gsfc.nasa.gov/docs/asca/.}.

\subsection{SIS}
\label{sis}

Most of the SIS data for these groups required corrections and
screening in addition to those performed under the standard Revision 2
Data Processing (REV2). Many of these observations were taken after
1994 at which time the effects of radiation damage on the SIS CCDs
began to become important (see Dotani et al. 1995). One of the
problems caused by radiation damage is the Residual Dark Distribution
(RDD) which is essentially an increase in the dark current of the CCDs
(see http://www.astro.isas.ac.jp/$\sim$dotani/rdd.html). RDD degrades
the energy resolution and the detection efficiency of the data.

Although in principle one can correct for the RDD effect, the
presently available RDD maps and software (i.e. FTOOL {\sc
correctrdd}) do not fully restore the SIS data for energies below
$\sim 0.8$ keV. The primary result is that the SIS data overestimate
the column density by $\sim 2\times 10^{20}$ cm$^{-2}$, an effect
which is more serious for the SIS1 (see the {\sl ASCA} GOF WWW
pages). This excess column is of similar magnitude to the Galactic
columns of the groups in our sample (see Table \ref{tab.obs}).

The most accurate correction for the RDD effect is possible for SIS
data taken in FAINT mode. After applying the RDD correction the FAINT
mode data is converted to BRIGHT2 mode data for which spectral
responses can be generated using the standard software.
Unfortunately, for several of the observations the exposures were
approximately evenly split between FAINT mode and BRIGHT mode. Since
the spectral responses of BRIGHT and BRIGHT2 mode data are different,
the data from these different modes cannot be combined. However, the
FAINT mode data instead can be converted to BRIGHT mode data (as is
done in the standard processing), and this BRIGHT mode data can be
corrected for the RDD effect.

Since the benefit from larger S/N outweighs the small improvement in
the RDD correction for FAINT mode data, for groups that require the
RDD correction but have a significant fraction of their exposure taken
in BRIGHT mode, we performed the RDD correction on the BRIGHT mode
data: i.e. both on the real BRIGHT data and the BRIGHT data that is
converted from FAINT mode. For other groups that require the RDD
correction, we applied the correction on the FAINT mode data which is
then converted to BRIGHT2 for further analysis. Finally, we analyzed
the BRIGHT mode data for all groups that do not require RDD
correction. (As is standard only data taken in medium and high bit
rates were used for the SIS.)

Most of the events files were filtered using the default REV2
screening criteria. The most frequent departure from the default
screening occurred because of temperature fluctuations in the SIS.
The data quality, most importantly the energy resolution, of the SIS
CCDs significantly degrades when the temperatures exceed $\sim -60
^{\circ}\rm C$, though optimal performance is achieved for
temperatures below $\sim -61 ^{\circ}\rm C$. In addition, the maps
required for the RDD correction are defined according to 1-degree
intervals in the SIS temperature; e.g. -$61 ^{\circ}\rm C$ to -$62
^{\circ}\rm C$. The current software allows the RDD correction to be
made with only one RDD map for each SIS. We defined maximum
temperatures considering these issues. In Table \ref{tab.obs} we list
the maximum temperatures for each SIS for those observations that
required temperature filtering.

We also improved the data quality for a few groups by using more
restrictive angles from the bright Earth (BR\_EARTH) and smaller
values for the Radiation Belt Monitor (RBM); these occurrences are
listed at the end of this section. Our final screening of the data
involved visual examination of the light curves for each observation
and removing intervals of anomalously high (or low) count rate. Note
only data with the standard event grades (0234) were used.

The gain of the SIS varies from chip-to-chip, and the variation is a
function of time owing to radiation damage. We have corrected the SIS
data for these effects using the most up-to-date calibration files as
of this writing (sisph2pi\_110397.fits).

The final processed events were then extracted from a region centered
on the emission peak for each detector of each sequence.  We selected
a particular extraction region using the following general
guidelines. Our primary concern is to select a region that encloses
most of the X-ray emission yet is symmetrically distributed about the
origin of the region; as a result, we used circles for most of our
extraction apertures.  We limited the size of the aperture to ensure
that the entire aperture fit on the detector, an issue more important
for the SIS because of its smaller field-of-view ($20\arcmin$ square
for SIS vs $50\arcmin$ diameter for GIS). Moreover, for the SIS0 and
SIS1 we tried to limit the apertures to as small a number of chips as
possible to reduce the effects of residual errors in chip-to-chip
calibration. 

We extracted the events using regions defined in detector coordinates
as is recommended in the {\it ASCA} ABC GUIDE because the spectral
response depends on the location in the detector not the position on
the sky. The radii for the SIS extraction regions are listed in Table
\ref{tab.obs}. (For NGC 1132 we used a square box for the SIS0 data and
a rectangular box for the SIS1 data. The radius listed for this group in
Table \ref{tab.obs} corresponds to one-half the length of the sides of
the SIS0 box. The SIS1 region is obtained by squashing the SIS0 box by
10\% along one of the coordinate axes.)

We emphasize that our SIS regions cover the area of significant S/N
and are negligibly contaminated by any obvious sources in the
field. Since the X-ray emission of these groups is quite extended
(e.g. Mulchaey \& Zabludoff 1998), our apertures typically enclose
$\sim 50$ per cent of the total emission. However, because the surface
brightness declines rapidly with radius, and our apertures enclose the
regions of significant S/N, the back-scattering of the emission at
larger radii into our apertures is neglible. In particular Takahashi
et al. \shortcite{takahashi} showed that approximately 94\% of the GIS
flux of a model cluster within an (on-axis) aperture $r=4\arcmin$
actually originates from within that aperture; an even larger
percentage applies for SIS data because the SIS+XRT PSF is smaller
than the GIS+XRT PSF.  Since the {\sl ASCA} XRT is smallest for
energies $\sim 0.5$-3 keV which are most important for groups, the
back-scattering of emission from larger radii into our apertures is at
most a few percent and thus has a negligible impact on our spectral
analysis.

Our apertures differ from those of Davis et al. \shortcite{dmm} who
instead defined most of their radii to encircle all of the emission
detected with {\sl ROSAT}. Their procedure leads to radii that are
factors of 5-10 larger than our values listed in Table
\ref{tab.obs}. However, the {\sl ASCA} data at such large radii are
very noisy and do not improve constraints on spectral models obtained
from using our smaller regions; i.e. we find that using larger
apertures increases the noise and generally loosens constraints on the
models obtained using data in the apertures enclosing the highest S/N
data.

For most of the observations we computed a background spectrum for
each detector using the standard deep observations of blank fields;
these are indicated by ``T'' under ``SIS BKG'' in Table \ref{tab.obs}.
There are important advantages to using these background templates
instead of a local background estimate. First, the templates allow
background to be extracted from the same parts of the detector as the
source and thus the vignetting and other exposure effects are the same
for each; this is not the case for background taken from a different
region of the detector as the current software does not allow the
required corrections to be made for spectral analysis. Second, as
stated above, it is known from {\sl ROSAT} observations of these
galaxies that their emission extends over much of the field of view of
the SIS and perhaps more (e.g. Mulchaey \& Zabludoff 1998).

However, it should be remembered that the background templates are
most appropriate for SIS data taken in 4-ccd mode early in the
mission. For more recent data taken in 1-cdd or 2-ccd mode, using
these background templates can lead to spurious effects particularly
for energies above $\sim 7$ keV where instrumental background
dominates the cosmic X-ray background. Systematic errors in the SIS at
these high energies are not so important for our study since the
constraints on spectral models are dominated by data at lower
energies. The integrated spectra of these sources within our apertures
are not dominated by background and, as a result, the deduced
temperatures and abundances are not overly affected whether the
templates or local background are used.

For those sources where the screening differed significantly from the
REV2 standard we used a local background estimate. The background
regions were chosen to be as far away as possible from the group
center and any other sources in the field. The observations using a
local background are listed with a ``L'' under ``SIS BKG'' in Table
\ref{tab.obs}. 

The final screened background-subtracted exposures and count rates are
listed in Table \ref{tab.exp}.

The instrument response matrix required for spectral analysis of {\sl
ASCA} data is the product of a spectral Redistribution Matrix File
(RMF) and an Auxiliary Response File (ARF). The RMF specifies the
channel probability distribution for a photon (i.e. energy resolution
information) while the ARF contains the information on the effective
area.  An RMF needs to be generated specifically for each SIS of each
observation because, among other reasons, each chip of each SIS
requires its own RMF and the spectral resolution of the SIS is
degrading with time. We generated the responses for each SIS using the
{\sc FTOOL sisrmg} selecting for the standard event grades
(0234). Using the response matrix and spectral PI (Pulse Invariant)
file we constructed an ARF file with the {\sc FTOOL ascaarf}.

The source apertures used for the SIS overlapped more than one chip
for NGC 5846 and NGC 6329. In order to analyze the spectra of such
regions we followed the standard procedure of creating a new response
matrix that is the average of the individual response matrices of each
chip weighted by the number of source counts of each chip (i.e. within
the source aperture). (Actually, the current software only allows the
RMFs to be averaged and then an ARF is generated using the averaged
RMF. This is considered to be a good approximation for most sources.)
Unfortunately, some energy resolution is lost as a result of this
averaging. For the observations of our sources, however, this small
energy broadening is rendered undetectable by the statistical noise of
the data.

\subsection{GIS}
\label{gis}

We analyzed the GIS events files processed by the standard REV2
screening. The only additional screening required for a few
observations involved removing time intervals with anomalously large
deviations from the mean light curve of the data.  No dead time
corrections were required for the GIS data because the count rates
were less than 1 ct s$^{-1}$ for each GIS detector for all the
observations in Table \ref{tab.exp}.

For consistency with the SIS data, the regions used to extract the GIS
data were chosen to be of similar size to the corresponding SIS
regions of a given sequence. Since, however, the GIS+XRT PSF is
somewhat larger than that of the SIS+XRT the extraction regions for
the GIS are usually $\sim 20\%$ larger. 

The standard observations of blank fields were used to estimate the
background for the GIS data. The principal motivation for using these
templates for GIS data is that the instrumental background increases
rapidly with increasing off-axis angle. We performed the rise-time
filtering (i.e. {\sc gisclean}) on the standard GIS templates as
required to match the standard screening. In Table \ref{tab.exp} the
background-subtracted exposures and count rates are listed for each
GIS observation.

The RMFs for the GIS2 and GIS3 are equivalent (and are assumed to be
independent of time), and thus for all GIS spectra we use the GIS RMFs
gis2v4\_0.rmf and its twin gis3v4\_0.rmf obtained from the HEASARC
archive. Using these RMF files an ARF file was generated for each
observation.

\subsection{Comments on individual groups}
\label{indiv}

The observations of several of the groups warrant further discussion.

\medskip

\noindent {\bf HCG 51:} The SIS were operated for $\sim 45$ ks in
2-ccd mode and for $\sim 27$ ks in 1-ccd mode. We separately reduced
the data for each mode and then combined the source and background
spectra, RMFs, and ARFs using the standard software (i.e. with FTOOL
{\sc addascaspec} v1.27).

\noindent {\bf MKW 9:} We adopted slightly stricter screening for the
SIS to help compensate for the degradation in data quality arising
from the warmer CCD temperatures: i.e. BR\_EARTH $>40$ and RBM $<50$.

\noindent {\bf NGC 533:} Each SIS for both the 1994 and 1996
observations were unfortunately quite warm for most of the
exposures. The maximum temperatures listed in Table \ref{tab.obs}
represent a compromise between a desire to (1) maximize data quality,
(2) maximize the acceptable exposure, and (3) to best match the
temperatures of the RDD maps available near the dates of the
observations. The maximum temperatures chosen significantly reduced
the effective exposure. As a result, since the maximum temperatures
still indicate that the data quality are not optimal, we followed the
advice in the {\sl ASCA} ABC guide and increased the acceptable number
of events by raising the threshold for pixel rejection to 100 for both
SIS of the 1996 observation; for the 1994 observation we raised the
threshold to 150 for the SIS0 data and to 200 for the SIS1 data.

Approximately 6 ks of 2-ccd mode data were also taken during
1994. These data were not analyzed since much of the group flux was
lost due to the unfortunate placement of the group center near the top
edge of the default chips SIS0 chip 1 and SIS1 chip3 (an issue more
serious for the SIS1 data). Finally, there is a noticeable point-like
source $\sim 7\arcmin$ from the group center. As a result, we limited
the sizes of our extraction regions, particularly the GIS because of
its larger PSF, to reduce contamination from this source.

\noindent {\bf NGC 2563:} Approximately $7\arcmin$ from the group
center there is a point-like source.

\noindent {\bf NGC 4104 (MKW 4s):} There were two observations taken
in 1996. One of the observations only yielded $\sim 3$ks of useful
exposure when restricting the maximum SIS temperatures to $-60
^{\circ}\rm C$ and thus was not included in our analysis.

\noindent {\bf NGC 4325:} The shape of the SIS0 spectrum for energies
$\sim 3$-5 keV is quite sensitive to the size of the region chosen. As
the region size is increased from our chosen $r=3\arcmin$ the spectrum
in that energy range flattens significantly. It is possible that an
unresolved hard source(s) contaminates the spectrum at larger
radii. More likely is that the SIS background template is inadequate
for this 1997 observation (see section \ref{sis}). To reduce the
sensitivity to either of these issues we choose the minimum region
size appropriate for the PSF. This choice is also justified because
(1) larger regions centered on the emission peak go off the edge of
the SIS, and (2) the emission is very centrally peaked.

\noindent {\bf NGC 5129:} There is a fairly bright point-like source
$\sim 7.5\arcmin$ from the group center.

\noindent {\bf NGC 5846:} We found that the standard REV2 events files
produce an SIS0 image having a noticeable excess stream of light on
the SIS0 chip 0. As a result, we screened the data ourselves and chose
slightly stricter screening criteria: i.e. BR\_EARTH $>40,30$ for the
SIS0, SIS1 and RBM $<50$ for both SIS.

\noindent {\bf NGC 6329:} The GIS2 data between 1-1.5 keV dip
significantly, an effect which is much less pronounced in the GIS3
data and not observed with the SIS data. We were unable to affect this
behavior in the GIS2 data by changing the screening criteria. Since
this dip is likely to be an instrumental effect we exclude the GIS2
data from analysis.

\noindent {\bf RGH 80:} There is a fairly bright source
$\sim 7\arcmin$ from the group center. We limit the sizes of the
regions to reduce contamination from this source.

\section{Broad-band spectral fitting}
\label{broad}

\subsection{Preliminaries}
\label{prelim}

\subsubsection{Models}
\label{models}

Since the X-ray emission of the groups in our sample is dominated by
hot gas, we use coronal plasma models as the basic component of our
spectral models. We use the MEKAL plasma code which is a modification
of the original MEKA code (Mewe, Gronenschild, \& van den Oord 1985;
Kaastra \& Mewe 1993) where the Fe L shell transitions crucial to the
X-ray emission of ellipticals and groups have been re-calculated
\cite{mekal}.  Although there exist other plasma codes, most notably
the popular Raymond-Smith code (1977, and updates), we have shown
previously (B99 and BF) that the MEKAL code provides a better
description of the {\sl ASCA} emission of the brightest elliptical
galaxies (which are similar to groups) than does the current version
of the Raymond-Smith code.

The superiority of the MEKAL code is expected because although both
the MEKAL and Raymond-Smith codes are identical in their treatment of
the ionization balance as given by Arnaud \& Raymond \shortcite{aray}
for Fe and Arnaud \& Rothenflug \shortcite{aroth} for the the other
elements, the Raymond-Smith code has many fewer lines than does MEKAL,
and, more importantly, it does not incorporate the improved
calculations of the Fe L shell lines by Liedahl et
al. \shortcite{mekal}. Despite these differences, qualitatively
similar results for ellipticals are usually obtained from the
different codes; we refer the reader to section 3.2.2 of B99 for a
detailed comparison. For our present investigation we shall employ
only the MEKAL code.

We account for absorption by our Galaxy using the photo-electric
absorption cross sections according to Baluci\'{n}ska-Church \&
McCammon \shortcite{phabs}. The absorber is modeled as a uniform
screen at zero redshift. The column density is generally allowed to be
a free parameter to indicate any additional absorption due to, e.g.,
intrinsic absorbing material, calibration errors, etc. Note, however,
that the column density obtained in this manner only estimates the
excess absorption because the actual value depends on the details of
the absorber; e.g., a patchy intrinsic absorber requires factors for
partial covering and the source redshift.

We take solar (photospheric) abundances according to Anders \&
Grevesse \shortcite{ag} which give an Fe abundance of $4.68\times
10^{-5}$ relative to H by number.  Since the meteoritic solar
abundances (Fe/H of $3.24\times 10^{-5}$ by number) are probably more
appropriate, the Fe abundances we quote should be raised by a factor
of 1.44 \cite{im}. We continue to use the photospheric abundances in
this paper to facilitate comparison with previous studies.

The simplest spectral model we consider is that of isothermal gas
which we refer to as the ``1T'' model. This model is expressed as $\rm
phabs\times MEKAL$, where ``phabs'' represents the absorption and
``MEKAL'' the thermal plasma emission. The free parameters of this
model are the column density, $N_{\rm H}$, the temperature, $T$, the
metallicity, $Z$, and the normalization (emission measure), $\rm
EM$. We focus on models where only the Fe abundance is allowed to be a
free parameter and the abundances of all other elements are
tied to Fe in their solar ratios; i.e. in this case $Z$ indicates the
Fe abundance. Although the Fe abundance is by far the most important
for spectral analysis of the groups in our sample, we also explore
models that allow the relative abundances of other elements to be
varied separately from Fe.

For the next level of complexity we consider models with two
temperatures (2T) which may be expressed as $\rm phabs_c\times MEKAL_c
+ phabs_h\times MEKAL_h$. For these ``two-phase'' models the first
component represents the ``colder'' gas and the second component the
``hotter'' gas. The temperatures and columns of each component are
free parameters. Similar to the 1T case, we focus on models where the
abundances of all the elements of a plasma component are tied to Fe in
their solar ratios; i.e. only the metallicity, $Z$, is a free
parameter.  However, we tie together the abundances of the hotter
component and those of the colder component. (In no case did we find
that relaxing this constraint improved the fits.) In addition we found
it useful to explore a model with three temperatures (3T) for some
galaxies which is just the obvious extension of the 2T case.

We also consider a multiphase cooling flow model (CF) which provides a
simple model for gas emitting over a range of temperatures
\cite{rjcool}.  This model assumes gas cools continuously at constant
pressure from some upper temperature, $T_{\rm max}$. The differential
emission measure is proportional to $\dot{M}/\Lambda(T)$, where
$\dot{M}$ is the mass deposition rate of gas cooling out of the flow,
and $\Lambda(T)$ is the cooling function of the gas (in our case, the
MEKAL plasma code). It should be emphasized that this is arguably the
simplest model of a cooling flow with mass drop-out. The advantage of
this particular model is that it is well studied, relatively easy to
compute, and a good fit to several ellipticals (e.g. BF and B99).

Since gas is assumed to be cooling from some upper temperature $T_{\rm
max}$, the cooling flow model requires that there be a reservoir of
gas emitting at temperature $T_{\rm max}$ but is not participating in
the cooling flow. To accommodate this scenario we define our standard
multiphase cooling flow model to have two components, $\rm
phabs_c\times CF + phabs_h\times MEKAL_h$, where the upper temperature
of the cooling flow, $T_{\rm max}$, and the temperature of the ambient
gas, $T_{\rm h}$, are tied together in the spectral fits. We denote
this model by ``CF+1T''; these CF+1T models thus have one less free
parameter than do the 2T models. We also consider the cases where (1)
only the CF component is present, and (2) a plasma component with
variable temperature is added.

Finally, for completeness we also consider adding a high-temperature
bremsstrahlung component (BREM) to these models to account for
possible emission from discrete sources in the central galaxies. Such
a component is not expected to be important for groups because of
their larger ratios of X-ray to optical luminosity with respect to
isolated ellipticals (e.g. Canizares et al. 1987).

\subsubsection{Calibration and fitting issues}
\label{issues}

The SIS provides substantially better constraints on the spectral
models than the GIS because the superior energy resolution and larger
effective area near 1 keV of the SIS allow better measurement of the
portion of the spectrum where the bright, temperature sensitive Fe L
shell lines dominate the X-ray emission of groups with average
temperatures around 1-1.5 keV (e.g. see section 3.1 of B99). In
addition, the SIS covers energies down to $\sim 0.5$ keV as opposed to
$\sim 0.8$ keV for the GIS. Since there exist calibration problems and
uncertainties with both the SIS and GIS (see {\sl ASCA} GOF WWW
pages), particularly for energies below 1 keV, we have modified the
spectra and fitting procedures accordingly.

As mentioned in section \ref{sis} radiation damage has reduced the
sensitivity of the SIS for energies below 1 keV and has become
increasingly important over the past two years. Consequently, for
observations taken before 1996 we set the minimum SIS energy to 0.55
keV and raise this to 0.65 keV for later observations. We chose this
date based on our previous analysis of the deviations in the SIS0 and
SIS1 spectra of the bright elliptical NGC 4636 observed in December
1995 (see Figure 5 of B99).

The GIS has been calibrated using observations of the Crab nebula
\cite{gis}. There are some reported problems with the gain of the GIS,
although the energy dependence of these problems have not yet been
reported (see {\sl ASCA} GOF WWW pages). Hence, we initially set the
lower energy limit for the GIS spectra to 0.8 keV after inspection of
the figures in Fukazawa et al. \shortcite{gis}. (Below we revise this
choice.)

We decided to analyze SIS and GIS data below 9 keV (similar to B99)
because at the highest energies the calibration errors associated with
the XRT become increasingly important \cite{gendreau}, and the
background level generally dominates the signal (thus amplifying any
errors in the background determination).

Errors in the calibration of the energy scale of the SIS and GIS
result in energy offsets of a few eV for the SIS and $\sim 50$ eV for
the GIS (see {\sl ASCA} GOF WWW pages). To compensate for these
systematic errors we follow B99 and allow the redshifts for all
detectors of the SIS and GIS to be free parameters. Note that in
practice first we determine the best fitting spectral model with the
redshifts of the detectors all fixed at their redshifts in Table
\ref{tab.prop}. Then the redshifts are allowed to be free to determine
if there is any final improvement in the fits.

All spectral fitting was performed with the software package {\sc
XSPEC} \cite{xspec}. We used the exact MEKAL models in {\sc XSPEC}
(i.e. model parameter ``switch'' set to 0), not those derived from
interpolating pre-computed tables.  For all fits we used the
${\chi^2}$ method implemented in its standard form in {\sc XSPEC}.

In order for the weights to be valid for the ${\chi^2}$ method the PI
bins for each source spectrum must be re-grouped so that each group
has at least 20 counts.  This is generally regarded as the standard
binning procedure. The background files also typically have at least
20 counts when their energy bins are grouped similarly to the source
spectra.  The background-subtracted count rate in a particular group
can be small, but the uncertainties in the source and background are
correctly propagated by {\sc XSPEC} to guarantee approximately
gaussian statistics for the statistical weights (K. Arnaud 1998,
private communication). To optimize S/N we further rebinned the
spectra into PI groups having a minimum of 40 and 80 counts
respectively for the SIS and GIS so that the energy groups do not
greatly over-sample the energy resolution of the individual
detectors. However, we emphasize that (like B99) we find that the
derived model parameters and the relative quality of the fits when
using this courser binning are nearly identical to those obtained
when using the standard binning.

Hence, with the SIS and GIS spectra so prepared initially we fitted
the models described in the previous section jointly to the available
SIS0, SIS1, GIS2, and GIS3 data for each group. By keeping the data of
these detectors separate we were able to assess the importance of
calibration differences between the various detectors. In particular,
we allowed the column densities and redshifts of each detector to be
fitted separately to allow for differences in the low energy responses
and the energy scales of the various detectors.

The procedure we follow for spectral fitting is as described in BF and
B99. That is, we begin by fitting a 1T model jointly to the spectra;
i.e. a single MEKAL model modified by Galactic absorption (see Table
\ref{tab.prop}). The free parameters are the temperature, metallicity,
and normalizations of each detector. We then examine whether allowing
$N_{\rm H}$ to be free significantly improves the fit. For the 2T
model we add another temperature component to the best-fitting 1T
model with the column density initially tied to the value of the first
component; the redshifts and abundances are also tied to those of the
first component. Once the new best 2T model is found we allow $N_{\rm
H}^{\rm h}$ to vary separately from $N_{\rm H}^{\rm c}$ and fit again
to obtain a new best fit. A similar procedure is followed for the
other multi-component models.

Upon fitting 1T models to the spectra of each group we found that the
column densities for each detector are consistent in essentially every
case within their $1\sigma$ errors; i.e. the statistical errors on the
derived $N_{\rm H}$ are larger than the systematic errors due to
calibration problems\footnote{Since the differences in $N_{\rm H}$
between the SIS0 and SIS1 are insignificant the less accurate RDD
correction applied to the BRIGHT mode data of some of the observations
is unimportant and thus is outweighed by the gain in exposure time
(see section \ref{sis}).}.  This same agreement between detectors is
found for $N_{\rm H}$ on both components of 2T and CF+1T models for
seven of the groups in our sample.

For the other five groups we found that $N_{\rm H}^c/N_{\rm H}^h\ga 1$
for the SIS data but $N_{\rm H}^c/N_{\rm H}^h\ll 1$ for the GIS
data. This occurs because the SIS data, which dominates the fits,
generally requires both model components to have temperatures near 1
keV. Because the GIS is not as sensitive to energies near 1 keV, the
GIS would rather have one component near 1 keV and the other component
with a much higher temperature. To accommodate this desire the GIS
raises the value of $N_{\rm H}^h$ to a very large value ($\sim (\rm
several)\times 10^{22}$ cm$^{-2}$) which suppresses the contribution
of the hotter temperature component below $\sim 2$ keV. Then the GIS
raises the emission measure of this hotter component to better fit the
higher energies.  (Note that the improvement in $\chi^2$ when allowing
the column densities of the different detectors to be varied
separately is always substantially less than the improvement of, e.g.,
a 2T model over a 1T model.)

Thus, in order to guarantee qualitative consistency between the
different components of the multi-temperature models it is necessary
to tie together the column densities between the different
detectors. This requirement is further justified by the agreement of
the column densities of the different detectors for the 1T
models. When tying $N_{\rm H}$ between different detectors the
measured value simply represents an average. Since the constraints on
such an average $N_{\rm H}$ provided by the GIS data are dwarfed by
the SIS data, and there exist differences in the calibration of the
GIS2 and GIS3 below 1 keV that are currently not understood (see
Figure 4 of Fukazawa et al. 1997), we raise the minimum energy of the
GIS and exclude energies below 1 keV as also done in B99. (We do not
notice any significant effect as a result of this choice.)

We also find no significant differences in the measured redshifts
between the SIS0 and SIS1 or the GIS2 and GIS3 for any of the groups;
i.e. only the redshift differences between the SIS and GIS are
important. Hence, it is sensible to further reduce the number of free
parameters and tie the together the redshifts of the SIS0 and SIS1 and
similarly the GIS2 and GIS3.

Since we intend to tie together $N_{\rm H}$ between the different
detectors as well as the SIS0/SIS1 and GIS2/GIS3 redshifts, we have
eliminated most of the value in analyzing separately the individual
detectors of the SIS and GIS. Since S/N is a premium for our analysis
we follow B99 and instead analyze the summed SIS=SIS0+SIS1 and summed
GIS=GIS2+GIS3 data. The small loss of information resulting from
averaging the SIS0 and SIS1 response matrices and the GIS2 and GIS3
ARF files is outweighed by the gain in S/N and the increase in the
speed of the spectral fitting afforded by the reduction in the number
of data sets. 

Since combining the data sets increases the number of counts per PI
bin, we slightly increase the PI binning criteria to a minimum of 50
counts for the summed SIS data and to 100 counts for the summed GIS
data. Note that similar to B99 we find that the best-fitting
parameters obtained from fitting the summed SIS and summed GIS data
differ insignificantly from those obtained when fitting all detectors
separately (with columns tied between the detectors).

Finally, we emphasize that obtaining the global minimum for the
multicomponent models often requires some effort (see BF and
B99). Often the fits land in a local minimum upon first adding another
component to a model. The typical case is illustrated as
follows. Starting with the best 1T model the 2T model is constructed
by adding a second temperature component with the abundance of the
second component tied to the first component and $T_{\rm h}$ set to
some value greater than 1 keV. Initially one finds that the
best-fitting abundance is small and similar to the 1T case while
$T_{\rm h}\ga 3$ keV.  By resetting the abundance to a large value
(e.g. 2 solar) and $T_{\rm h}$ to a smaller value near 1 keV a new
deeper minimum is generally obtained.

For example, the initial 2T fit to NGC 5129 gives a value of
$\chi^2=41.3$ (32 dof) which is an improvement over the 1T value of
47.8 (36 dof). The best fitting abundance is $0.14Z_{\sun}$ which is
similar to the 1T value of $0.13Z_{\sun}$, and the best-fitting
$T_{\rm h}$ is 100 keV which is the maximum allowed by the MEKAL
model. However, upon resetting $Z$ and $T_{\rm h}$ following the above
prescription the fit improves further to $\chi^2=36.1$ and gives
best-fitting values $Z=0.69Z_{\sun}$ and $T_{\rm h}=1.46$ keV which
are significantly different from the initial fit (see section
\ref{results}).

Therefore, after a fit is completed we always reset a subset of the
free parameters (especially $Z$ and $T_{\rm h}$) and fit again several
times until we are satisfied that the minimum is stable. Stepping
through the parameters when determining confidence limits also was
useful in assessing the stability of the minimum. This is not the most
rigorous method to find the global minimum, but it is at present the
most convenient way to do it in {\sc XSPEC}.

\subsection{Results}
\label{results}

\begin{table*}
\caption{Spectral Fits}
\label{tab.fits}
\begin{tabular}{ccccccccrrc}
Group & $N_{\rm H}^{\rm c}$ & $N_{\rm H}^{\rm h}$ & $T_{\rm c}$ &
$T_{\rm h}$ & $Z$ & $\rm EM_c$ & $\rm EM_h$ & $\chi^2$ & dof & $P$\\  
     & ($10^{21}$ cm$^{-2}$) & ($10^{21}$ cm$^{-2}$) & (keV) & (keV) &
$(Z_{\sun})$ & \multicolumn{2}{c}{(see notes)}\\ 
HCG 51:\\
1T & $0.69_{-0.20}^{+0.22}$ & $\cdots$ & $1.35_{-0.05}^{+0.05}$ &
$\cdots$ & $0.33_{-0.06}^{+0.07}$ & $1.74_{-0.16}^{+0.18}$ & $\cdots$
& 158.3 & 115 & 4.6e-3\\
2T & $3.17_{-3.17}^{+1.95}$ & $0.77_{-0.41}^{+0.51}$ &
$0.68_{-0.13}^{+0.15}$ & $1.49_{-0.09}^{+0.12}$ &
$0.54_{-0.12}^{+0.18}$ & $0.34_{-0.26}^{+0.42}$ &
$1.23_{-0.18}^{+0.27}$ & 137.1 & 111 & 4.7e-2\\
CF+1T & $2.92_{-2.09}^{+0.94}$ & $0.59_{-0.59}^{+0.99}$ &
$1.54_{-0.12}^{+0.17}$ & tied & $0.58_{-0.16}^{+0.24}$ &
$7.15_{-4.36}^{+4.09}$ & $0.86_{-0.37}^{+0.43}$ & 138.1 & 112 &
4.8e-2\\ \\[-5 pt]
HCG 62:\\
1T & $0.00_{-0.00}^{+0.27}$ & $\cdots$ & $1.07_{-0.04}^{+0.04}$ &
$\cdots$ & $0.21_{-0.04}^{+0.04}$ & $3.48_{-0.31}^{+0.41}$ & $\cdots$
& 123.9 & 61 & 3.5e-6\\
2T & $1.32_{-1.32}^{+1.36}$ & $1.44_{-1.44}^{+1.41}$ &
$0.69_{-0.05}^{+0.06}$ & $1.44_{-0.16}^{+0.27}$ &
$0.99_{-0.38}^{+0.94}$ & $0.69_{-0.38}^{+0.59}$ &
$1.24_{-0.45}^{+0.50}$ & 63.3 & 57 & 0.26 \\
CF & $1.78_{-0.49}^{+0.41}$ & $\cdots$ & $1.69_{-0.12}^{+0.18}$ & $\cdots$
& $0.82_{-0.22}^{+0.39}$ & $7.90_{-1.77}^{+1.72}$ & $\cdots$ & 70.8
& 61 & 0.18 \\ \\[-5 pt]
MKW 9:\\
1T & $1.38_{-0.41}^{+0.44}$ & $\cdots$ & $2.16_{-0.14}^{+0.15}$ &
$\cdots$ & $0.45_{-0.12}^{+0.15}$ & $2.35_{-0.24}^{+0.27}$ & $\cdots$
 & 110.0 & 92 & 9.8e-2\\
2T & $4.72_{-2.97}^{+1.98}$ & $3.14_{-1.23}^{+1.26}$ &
$0.65_{-0.12}^{+0.15}$ & $2.30_{-0.22}^{+0.27}$ &
$0.70_{-0.22}^{+0.31}$ & $0.56_{-0.39}^{+0.68}$ &
$2.15_{-0.34}^{+0.39}$ & 89.7 & 88 & 0.43\\
CF+1T & $2.28_{-2.28}^{+1.64}$ & $2.81_{-1.54}^{+1.91}$ &
$2.42_{-0.33}^{+0.59}$ & tied & $0.57_{-0.17}^{+0.24}$ &
$8.29_{-6.51}^{+8.77}$ & $1.87_{-0.90}^{+0.65}$ & 94.2 & 89 & 0.33\\
\\[-5 pt] 
NGC 533:\\
1T & $0.00_{-0.00}^{+0.60}$ & $\cdots$ & $1.26_{-0.09}^{+0.06}$ &
$\cdots$ & $0.27_{-0.07}^{+0.09}$ & $1.95_{-0.25}^{+0.38}$ & $\cdots$
& 90.1 & 64 & 1.8e-2\\
2T & $2.94_{-2.94}^{+2.16}$ & $2.82_{-2.04}^{+1.34}$ &
$0.63_{-0.10}^{+0.09}$ & $1.37_{-0.18}^{+0.29}$ &
$1.16_{-0.67}^{+6.84}$ & $0.46_{-0.39}^{+0.81}$ &
$0.90_{-0.66}^{+0.80}$ & 68.5 & 58 & 0.16\\
CF+1T & $2.77_{-2.77}^{+0.84}$ & $0.73_{-0.73}^{+3.42}$ &
$1.56_{-0.31}^{+0.19}$ & tied & $0.80_{-0.36}^{+0.88}$ &
$7.31_{-2.59}^{+3.55}$ & $0.38_{-0.32}^{+1.08}$ & 73.5 & 59 &
9.7e-2\\ \\[-5 pt] 
NGC 1132:\\
1T & $0.02_{-0.02}^{+0.45}$ & $\cdots$ & $1.15_{-0.05}^{+0.04}$ &
$\cdots$ & $0.26_{-0.05}^{+0.06}$ & $2.20_{-0.22}^{+0.30}$ & $\cdots$
& 71.2 & 69 & 0.40\\
2T & $3.87_{-3.10}^{+2.07}$ & $0.00_{-0.00}^{+0.63}$ &
$0.71_{-0.18}^{+0.19}$ & $1.31_{-0.11}^{+0.17}$ &
$0.57_{-0.22}^{+0.41}$ & $0.92_{-0.66}^{+0.93}$ &
$1.18_{-0.43}^{+0.63}$ & 67.7 & 65 & 0.39\\
CF+1T & $3.40_{-3.40}^{+2.21}$ & $0.00_{-0.00}^{+1.40}$ &
$1.22_{-0.09}^{+0.14}$ & tied & $0.43_{-0.17}^{+0.40}$ &
$13.05_{-12.15}^{+8.70}$ & $1.10_{-0.66}^{+0.99}$ & 68.9 & 66 &
0.38\\ \\[-5 pt] 
NGC 2563:\\
1T & $0.85_{-0.41}^{+0.47}$ & $\cdots$ & $1.35_{-0.10}^{+0.11}$ &
$\cdots$ & $0.36_{-0.11}^{+0.15}$ & $0.76_{-0.14}^{+0.16}$ & $\cdots$
& 68.0 & 60 & 0.22\\
2T & $4.36_{-4.36}^{+2.03}$ & $0.40_{-0.40}^{+2.18}$ &
$0.80_{-0.13}^{+0.18}$ & $1.81_{-0.36}^{+0.49}$ &
$0.91_{-0.45}^{+1.01}$ & $0.29_{-0.23}^{+0.34}$ &
$0.30_{-0.13}^{+0.31}$ & 52.5 & 56 & 0.61\\
CF+1T & $3.37_{-2.19}^{+0.99}$ & $0.00_{-0.00}^{+4.97}$ &
$1.81_{-0.30}^{+0.52}$ & tied & $0.90_{-0.38}^{+0.67}$ &
$2.12_{-1.28}^{+0.94}$ & $0.10_{-0.10}^{+0.39}$ & 53.5 & 57 & 0.61\\
\\[-5 pt] 
NGC 4104:\\
1T & $0.12_{-0.12}^{+0.29}$ & $\cdots$ & $2.15_{-0.13}^{+0.13}$ &
$\cdots$ & $0.57_{-0.14}^{+0.16}$ & $1.43_{-0.15}^{+0.17}$ & $\cdots$
& 131.7 & 109 & 6.9e-2\\
2T & $2.73_{-2.73}^{+2.37}$ & $1.12_{-0.98}^{+1.17}$ &
$0.76_{-0.12}^{+0.23}$ & $2.30_{-0.26}^{+0.38}$ &
$0.70_{-0.25}^{+0.32}$ & $0.32_{-0.21}^{+0.38}$ &
$1.11_{-0.22}^{+0.26}$ & 109.7 & 103 & 0.31\\
CF+1T & $2.23_{-0.54}^{+0.53}$ & $0.00_{-0.00}^{+3.17}$ &
$3.10_{-0.40}^{+0.32}$ & tied & $0.71_{-0.20}^{+0.26}$ &
$5.54_{-1.55}^{+1.89}$ & $0.35_{-0.25}^{+0.29}$ & 114.2 & 104 &
0.23\\ \\[-5 pt] 
NGC 4325:\\
1T & $0.96_{-0.43}^{+0.52}$ & $\cdots$ & $0.85_{-0.04}^{+0.04}$ &
$\cdots$ & $0.21_{-0.03}^{+0.04}$ & $5.29_{-0.53}^{+0.61}$ & $\cdots$
& 112.8 & 89 & 4.5e-2\\ 
2T & $3.17_{-0.86}^{+0.80}$ & $0.22_{-0.22}^{+0.92}$ &
$0.66_{-0.07}^{+0.05}$ & $1.09_{-0.11}^{+0.14}$ &
$0.44_{-0.11}^{+0.20}$ & $3.58_{-1.14}^{+1.12}$ &
$1.27_{-0.41}^{+0.83}$ & 83.6 & 85 & 0.52\\ 
CF+1T & $3.42_{-0.67}^{+0.52}$ & $1.42_{-1.07}^{+1.08}$ &
$0.97_{-0.07}^{+0.12}$ & tied & $0.48_{-0.15}^{+0.28}$ &
$74.53_{-22.56}^{+21.22}$ & $1.13_{-0.93}^{+1.04}$ & 88.6 & 86 &
0.40\\ \\[-5 pt] 
NGC 5129:\\
1T & $0.00_{-0.00}^{+1.11}$ & $\cdots$ & $0.92_{-0.10}^{+0.05}$ &
$\cdots$ &  $0.13_{-0.03}^{+0.04}$  & $1.05_{-0.15}^{+0.32}$ &
$\cdots$ & 47.8 & 36 & 9.0e-2\\ 
2T & $2.62_{-2.61}^{+1.48}$ & $0.00_{-0.00}^{+4.21}$ &
$0.68_{-0.12}^{+0.10}$ & $1.46_{-0.40}^{+0.88}$ &
$0.69_{-0.41}^{+4.10}$ & $0.42_{-0.32}^{+0.53}$ &
$0.24_{-0.24}^{+0.34}$ & 36.1 & 32 & 0.28\\
CF & $2.29_{-0.98}^{+1.48}$ & $\cdots$ &  $1.37_{-0.28}^{+0.27}$ &
$\cdots$ &  $0.58_{-0.25}^{+0.58}$ & $7.96_{-3.04}^{+7.76}$ & $\cdots$
& 37.9 & 36 & 0.38\\ \\[-5 pt]
NGC 5846:\\
1T & $1.79_{-0.58}^{+0.74}$ & $\cdots$ & $0.68_{-0.04}^{+0.04}$ &
$\cdots$ & $0.26_{-0.06}^{+0.12}$ & $5.59_{-0.67}^{+0.77}$ & $\cdots$
& 178.0 & 79 & 1.3e-9\\ 
2T & $2.69_{-0.64}^{+0.57}$ & $0.00_{-0.00}^{+0.79}$ &
$0.61_{-0.03}^{+0.03}$ & $2.52_{-0.54}^{+1.09}$ &
$1.15_{-0.44}^{+1.12}$ & $1.87_{-0.81}^{+0.98}$ &
$0.41_{-0.12}^{+0.15}$ & 98.8 & 75 & 3.4e-2\\
CF & $1.88_{-0.37}^{+0.64}$ & $\cdots$ & $1.11_{-0.13}^{+0.09}$ &
$\cdots$ & $0.43_{-0.11}^{+0.14}$ & $3.28_{-0.65}^{+1.56}$ &
$\cdots$ & 137.7 & 79 & 4.9e-5\\ \\[-5 pt]
NGC 6329:\\
1T & $0.00_{-0.00}^{+0.28}$  & $\cdots$ & $1.40_{-0.10}^{+0.11}$ &
$\cdots$ & $0.21_{-0.07}^{+0.10}$ & $1.00_{-0.13}^{+0.15}$ & $\cdots$
& 34.8 & 38 & 0.62\\
2T & $0.42_{-0.42}^{+1.45}$ & tied & $0.72_{-0.17}^{+0.30}$ &
$1.69_{-0.21}^{+0.29}$ & $0.53_{-0.24}^{+0.55}$ &
$0.10_{-0.05}^{+0.12}$ & $0.68_{-0.18}^{+0.19}$ & 23.4 & 35 & 0.93\\
CF & $1.72_{-0.68}^{+0.69}$ & $\cdots$ & $2.41_{-0.36}^{+0.46}$ &
$\cdots$ & $0.58_{-0.24}^{+0.43}$ & $5.15_{-1.49}^{+2.04}$ & $\cdots$
& 26.2 & 38 & 0.93\\ \\[-5 pt]
RGH 80:\\
1T & $0.19_{-0.19}^{+0.33}$ & $\cdots$ & $1.20_{-0.06}^{+0.06}$ &
$\cdots$ & $0.26_{-0.05}^{+0.06}$ & $1.77_{-0.22}^{+0.26}$ & $\cdots$
& 125.5 & 69 & 3.8e-5\\
2T & $2.42_{-1.22}^{+1.04}$ & $0.00_{-0.00}^{+0.42}$ &
$0.78_{-0.06}^{+0.07}$ & $1.64_{-0.17}^{+0.21}$ &
$0.67_{-0.24}^{+0.42}$ & $0.68_{-0.25}^{+0.37}$ &
$0.61_{-0.16}^{+0.18}$ & 79.8 & 65 & 0.10\\
CF & $2.03_{-0.44}^{+0.44}$ & $\cdots$ & $1.86_{-0.08}^{+0.19}$ &
$\cdots$ & $0.89_{-0.26}^{+0.38}$ & $23.87_{-4.94}^{+5.81}$ & $\cdots$
& 102.3 & 69 & 5.7e-3\\
\end{tabular}
\smallskip

\raggedright

Best-fitting values and 90\% confidence limits on one interesting
parameter $(\Delta\chi^2=2.71)$ are listed for isothermal (1T),
two-temperature (2T), and multiphase cooling flow models (CF or CF+1T)
jointly fit to the SIS and GIS data; the values of $\chi^2$, the
number of degrees of freedom (dof), and the $\chi^2$ null hypothesis
probability $(P)$ are also given. Abundances are quoted in units of
their (photospheric) solar values \cite{ag}; i.e. Fe abundance is
$4.68\times 10^{-5}$ relative to H. For models with multiple
temperature components the abundances of the hotter component are tied
to the corresponding abundances of the colder component during the
fits.  For CF+1T models $T_{\rm h}$ is tied to $T_{\rm c}$ during the
fits. The emission measures $(\rm EM)$ are quoted in units of
$10^{-17}n_en_pV/4\pi D^2$ similar to what is done in {\sc XSPEC}
except for the CF components where instead $\rm EM_c$ refers to
$\dot{M}$ in solar masses per year using the distances described in
the notes of Table \ref{tab.prop}. See section \ref{models} for
further explanation of the models.

\end{table*}

\begin{table*}
\caption{Three-Component Spectral Fits for NGC 5846}
\label{tab.n5846}
\begin{tabular}{ccccccccc}
$N_{\rm H}^{\rm 1}$ & $N_{\rm H}^{\rm 2}$ & $T_{\rm 1}$ &
$T_{\rm 2}$ & $T_{\rm 3}$ & $Z$\\   
($10^{21}$ cm$^{-2}$) & ($10^{21}$ cm$^{-2}$) & (keV) & (keV) &
(keV) & $(Z_{\sun})$ & $\rm EM_1$ & $\rm EM_2$ & $\rm EM_3$ \\ \\[-5pt] 
\multicolumn{9}{c}{\normalsize 3T: ($\chi^2=81.0$, 72 dof,
$P=0.22$)}\\ \\[-7pt] 
$0.48_{-0.48}^{+2.42}$ & $4.42_{-0.92}^{+1.30}$ &
$0.50_{-0.18}^{+0.09}$ & $0.76_{-0.08}^{+0.10}$ &
$3.27_{-0.88}^{+2.86}$ & $3.83_{-2.53}^{+\infty}$ &
$0.14_{-0.13}^{+0.52}$ & $0.55_{-0.48}^{+0.82}$ &
$0.16_{-0.06}^{+0.13}$\\ \\
\multicolumn{9}{c}{\normalsize CF+2T: ($\chi^2=82.7$, 73 dof,
$P=0.20$)}\\ \\[-7pt] 
$2.23_{-1.92}^{+0.89}$ & $5.72_{-2.04}^{+8.24}$ &
$0.77_{-0.06}^{+0.11}$ & tied & $3.72_{-0.98}^{+\infty}$ &
$2.00_{-0.92}$ & $5.70_{-4.22}^{+3.52}$ & $0.66_{-0.44}^{+0.87}$ &
$0.20_{-0.11}^{+0.11}$\\  
\end{tabular}
\medskip

\raggedright

Best-fitting values and 90\% confidence limits on one interesting
parameter $(\Delta\chi^2=2.71)$ are listed for three-component
models jointly fit to the SIS and GIS data of NGC 5846; see Table
\ref{tab.fits} for additional explanation. The column density of the
third component is tied to $N_{\rm H}^{\rm 1}$ for the 3T model and
is fixed to the Galactic value for the CF+2T model. For the cooling
flow model (1) $T_2$ is tied to the upper temperature of the cooling
flow ($T_1$) and (2) $\rm EM_1$ is $\dot{M}$. 

\end{table*}

\begin{figure*}
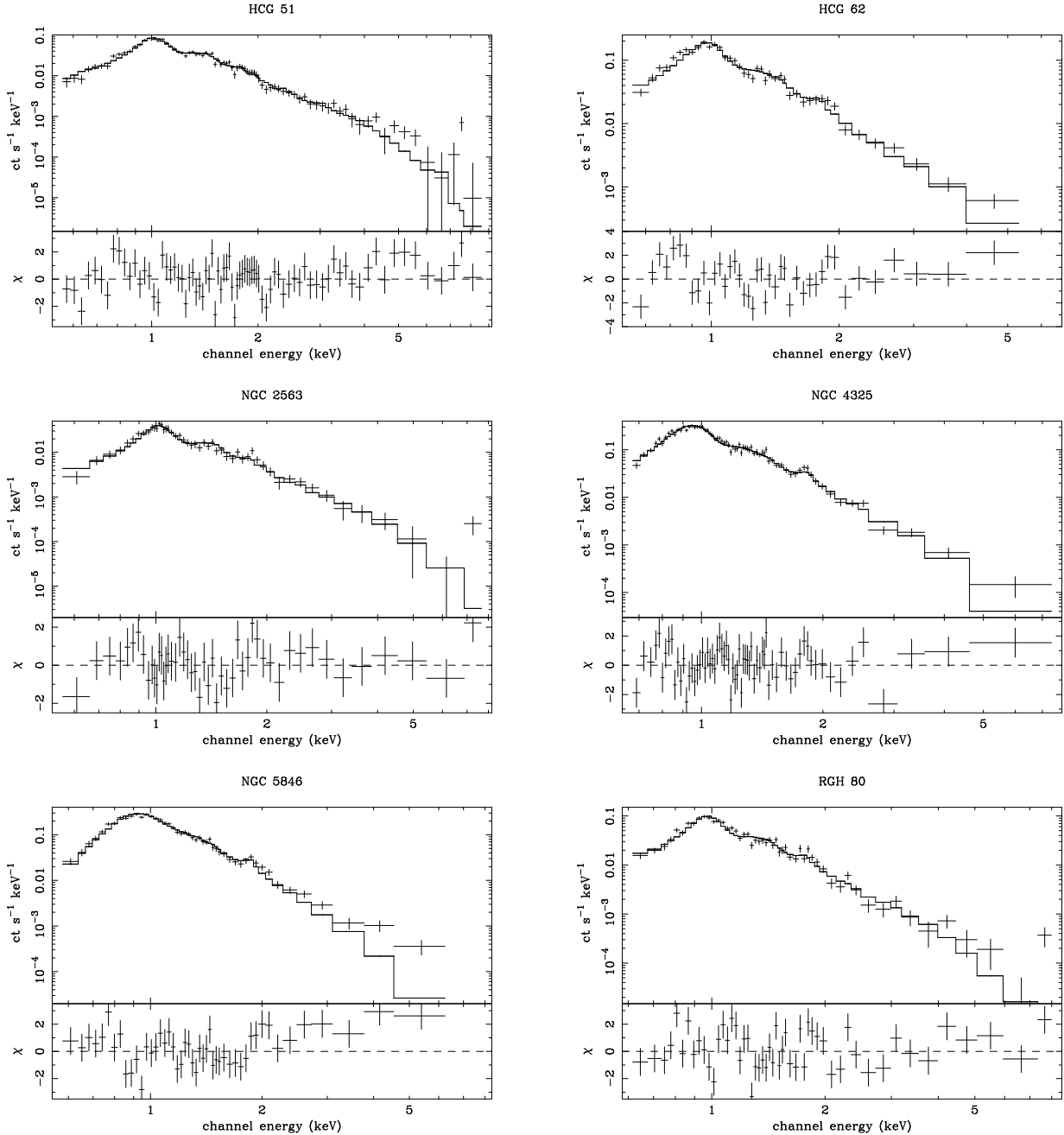

\parbox{0.49\textwidth}{
\centerline{\psfig{figure=h51_1t.ps,angle=-90,height=0.23\textheight}}
}
\parbox{0.49\textwidth}{
\centerline{\psfig{figure=h62_1t.ps,angle=-90,height=0.23\textheight}}
}

\vskip 0.4cm

\parbox{0.49\textwidth}{
\centerline{\psfig{figure=n2563_1t.ps,angle=-90,height=0.23\textheight}}
}
\parbox{0.49\textwidth}{
\centerline{\psfig{figure=n4325_1t.ps,angle=-90,height=0.23\textheight}}
}

\vskip 0.4cm

\parbox{0.49\textwidth}{
\centerline{\psfig{figure=n5846_1t.ps,angle=-90,height=0.23\textheight}}
}
\parbox{0.49\textwidth}{
\centerline{\psfig{figure=r80_1t.ps,angle=-90,height=0.23\textheight}}
}

\caption{\label{fig.1t} Best-fitting isothermal (1T) models fitted
jointly to the SIS and GIS data for six groups with among the best S/N
data in our sample. These models correspond to the 1T results listed
in Table \ref{tab.fits}. Only the SIS data are shown.}

\end{figure*}

\begin{figure*}
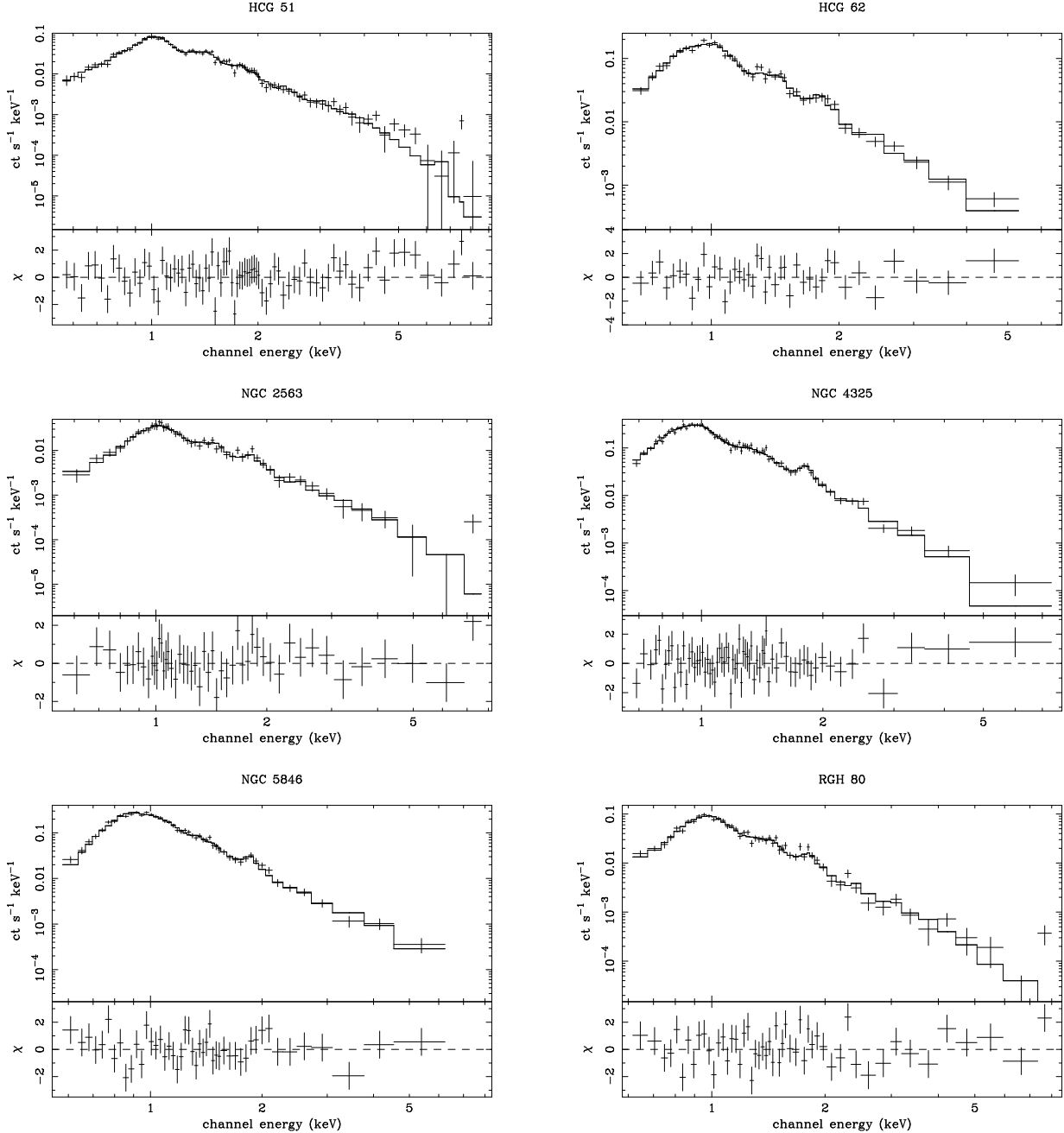

\parbox{0.49\textwidth}{
\centerline{\psfig{figure=h51_2t.ps,angle=-90,height=0.23\textheight}}
}
\parbox{0.49\textwidth}{
\centerline{\psfig{figure=h62_2t.ps,angle=-90,height=0.23\textheight}}
}

\vskip 0.4cm

\parbox{0.49\textwidth}{
\centerline{\psfig{figure=n2563_2t.ps,angle=-90,height=0.23\textheight}}
}
\parbox{0.49\textwidth}{
\centerline{\psfig{figure=n4325_2t.ps,angle=-90,height=0.23\textheight}}
}

\vskip 0.4cm

\parbox{0.49\textwidth}{
\centerline{\psfig{figure=n5846_2t.ps,angle=-90,height=0.23\textheight}}
}
\parbox{0.49\textwidth}{
\centerline{\psfig{figure=r80_2t.ps,angle=-90,height=0.23\textheight}}
}

\caption{\label{fig.2t} As Figure \ref{fig.1t} but for two-temperature
(2T) models.}

\end{figure*}

\begin{figure*}
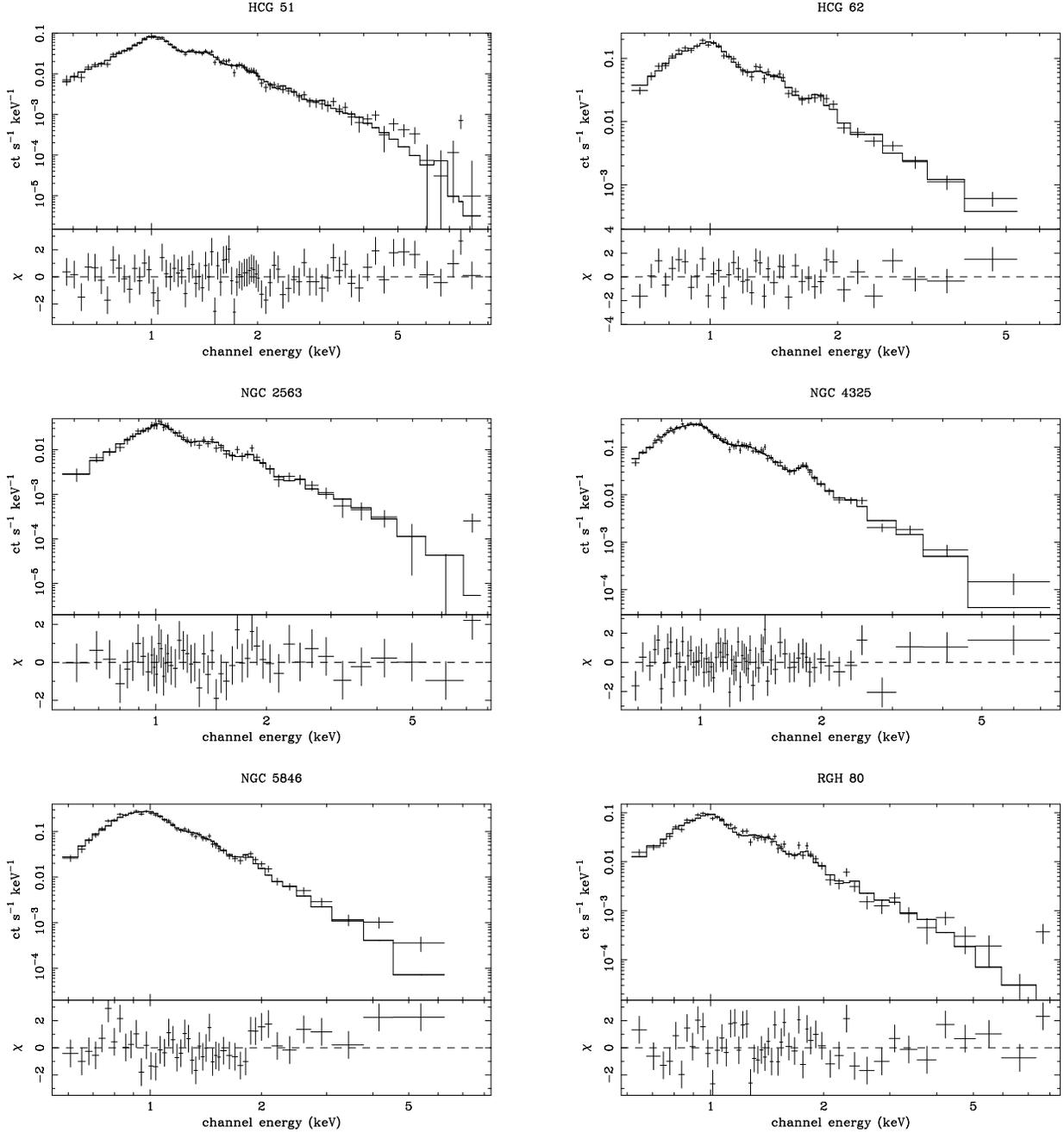

\parbox{0.49\textwidth}{
\centerline{\psfig{figure=h51_cf.ps,angle=-90,height=0.23\textheight}}
}
\parbox{0.49\textwidth}{
\centerline{\psfig{figure=h62_cf.ps,angle=-90,height=0.23\textheight}}
}

\vskip 0.4cm

\parbox{0.49\textwidth}{
\centerline{\psfig{figure=n2563_cf.ps,angle=-90,height=0.23\textheight}}
}
\parbox{0.49\textwidth}{
\centerline{\psfig{figure=n4325_cf.ps,angle=-90,height=0.23\textheight}}
}

\vskip 0.4cm

\parbox{0.49\textwidth}{
\centerline{\psfig{figure=n5846_cf.ps,angle=-90,height=0.23\textheight}}
}
\parbox{0.49\textwidth}{
\centerline{\psfig{figure=r80_cf.ps,angle=-90,height=0.23\textheight}}
}

\caption{\label{fig.cf} As Figure \ref{fig.1t} but for multiphase
cooling flow (CF or CF+1T) models.}

\end{figure*}

\begin{figure*}
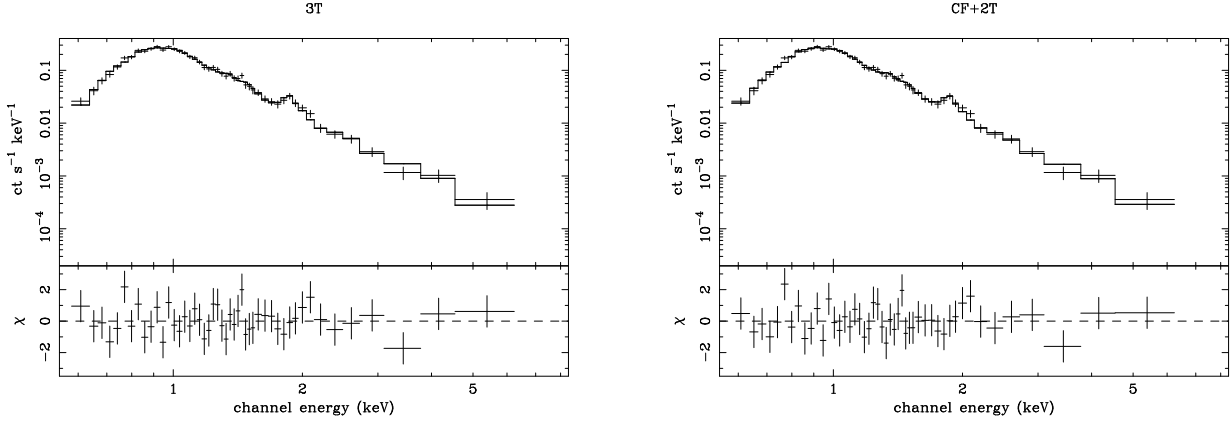

\parbox{0.49\textwidth}{
\centerline{\psfig{figure=n5846_3t.ps,angle=-90,height=0.23\textheight}}
}
\parbox{0.49\textwidth}{
\centerline{\psfig{figure=n5846_cf2t.ps,angle=-90,height=0.23\textheight}}
}
\caption{\label{fig.n5846} Best-fitting three-component (3T and CF+2T)
models for NGC 5846 corresponding to those listed in Table
\ref{tab.n5846}.  Although the models were fitted jointly to both the
SIS and GIS data, only the SIS data are shown.}

\end{figure*}

In sum: for each galaxy group we analyze the SIS data which are a sum
of the SIS0 and SIS1 data, and similarly the GIS data which are a sum
of the GIS2 and GIS3 data. The SIS data are restricted to 0.55-9 keV
for observations before 1996 and to 0.65-9 keV for observations since
1996. The energy range for the GIS data is 1-9 keV for all
observations. Finally, the PI bins are regrouped so that each group
has a minimum of 50 counts for the SIS and 100 counts for the GIS.

Table \ref{tab.fits} lists the results of fitting 1T, 2T, and cooling
flow (CF and CF+1T) models jointly to the SIS and GIS data of each
group. Comments on individual groups are given in section
\ref{comments}. 

\subsubsection{Isothermal models (1T)}
\label{1t}

In Figure \ref{fig.1t} we plot the best-fitting 1T models and
residuals corresponding to the results in Table \ref{tab.fits} for six
of the groups possessing among the best S/N data in our sample.  Only
the SIS data is shown because of space considerations and since the
GIS residuals are not nearly so prominent to the eye owing to the
poorer energy resolution of the GIS. (We refer the interested reader
to B99 for similar plots of the bright elliptical NGC 1399 which show
GIS data and residuals.)

In terms of the $\chi^2$ null hypothesis probability, $P$, we find
that for most of the groups the 1T models yield fits that are of poor
quality or marginal quality ($P\la 0.1$). Inspection of Figure
\ref{fig.1t} shows that the poor fits are largely the result of the
inability to fit the SIS data near 1 keV. These groups all display
similar residuals near 1 keV which are analogous to those found for
the brightest elliptical galaxies by BF and B99. Notice that NGC 2563
has prominent 1 keV residuals even though $P=0.22$.

Some of the groups have significant residuals at higher energies such
as HCG 51, NGC 4325, and NGC 5846. The residual pattern above $\sim 2$
keV for NGC 5846 clearly indicates emission from a higher temperature
component (see section \ref{2t}). However, for the other systems the
residuals at higher energies probably indicate errors in the
background subtraction (see section \ref{comments}). We mention that
such remaining errors in the background are not very important for our
current models because the results do not change appreciably if the
energies above $\sim 3$ keV are excluded from the fits; i.e. the SIS
data near 1 keV dominate the fits.

Our conclusion that the 1T models are generally poor fits to the {\sl
ASCA} spectra of these groups is qualitatively consistent with the
$\chi^2$ values (and their implied $P$-values) listed in Table 2 of
Davis et al. \shortcite{dmm} for several of the groups. (The poor
quality of the 1T fits is not discussed by Davis et al.) For other
groups, especially HCG 62, NGC 4104, and RGH 80, we obtain
substantially smaller $P$ values. The ability of our spectra to better
discriminate between models is not surprising because our spectra are
derived from much smaller spatial apertures where the S/N is optimal
and contamination from unresolved sources is significantly less
important. 

Most of the fitted column densities are consistent with the Galactic
values (Table \ref{tab.prop}) when considering the SIS calibration
uncertainties and, more importantly, the statistical errors. Only MKW
9 and especially NGC 5846 show significant excess absorption, although
interpretation in these cases is difficult because the fits are of
poor quality. In contrast, some galaxies have best-fitting column
densities {\sl below} the Galactic value which are only marginally
consistent within the estimated 90\% errors with the Galactic values
(e.g. HCG 62, NGC 1132, NGC 6329). For example, in Figure \ref{fig.1t}
the 1T model for energies less than 1 keV lies clearly below the data
for HCG 62 even with the best-fitting $N_{\rm H}=0$. Considering that
the SIS calibration errors should translate to a measured excess
$N_{\rm H}$ such examples give additional evidence against the
suitability of the 1T model.

The metallicity derived for most of the groups is very sub-solar: for
the entire sample we obtain a mean and standard deviation $\langle
Z\rangle = 0.29\pm 0.12$~$Z_{\sun}$. These low abundances are
consistent with the previous study of these groups by Davis et
al. \shortcite{dmm} who fitted 1T models to the {\sl ASCA}
spectra. Similar low Fe abundances are obtained by Fukazawa et
al. \shortcite{fuk96} for HGC 51, by Finoguenov et
al. \shortcite{fino} for NGC 5846, and by Finoguenov \& Ponman
\shortcite{fp} for HCG 51 and HCG 62 within our apertures. Fukazawa et
al. \shortcite{fuk98} found similar low Fe abundances for HCG 51, HCG
62, MKW 9, and NGC 4104 (MKW 4s), although their analysis excluded
most of the emission at small radii which dominates our
apertures. Such sub-solar abundances are also consistent with 1T
models fitted to elliptical galaxies (e.g. Matsumoto et al. 1997; BF;
B99). However, as we show below in sections \ref{2t}, \ref{cf}, and
Appendix \ref{bias}, in most cases these low abundances result from a
bias inherent in 1T models.

We have examined models where the $\alpha$-process elements O, Ne, Mg,
Si, and S are allowed to be varied separately from Fe. Generally the
largest improvement in the fits arises from letting the O and Ne
abundances go to zero while lesser improvements result when the Si and
S abundances take values $\sim 1.5$-2 $Z_{\rm Fe}$. For most of the
groups the improvements in the quality of the fits are insignificant,
and for those cases with noticeable increase in $P$ the improvement is
far less than when adding another temperature component as we discuss
below.

Similar Si and S enhancements are reported by Fukazawa et
al. \shortcite{fuk98}, Davis et al. \shortcite{dmm}, and Finoguenov \&
Ponman \shortcite{fp}. Since (1) the $\alpha$/Fe ratios are generally
poorly constrained and are consistent with solar within the 90\%
uncertainties, (2) the $\alpha$/Fe ratios, like the Fe abundance, turn
out to be highly model dependent (see below and Appendix \ref{bias}),
and (3) the 1T models with variable $\alpha$/Fe ratios are in any
event still inadequate fits to the data, we do not present detailed
results of these models. We refer the interested reader to B99 for
detailed exploration of such models for four bright ellipticals for
which results entirely analogous to the groups in our sample are
obtained.

\subsubsection{Two-temperature models (2T)}
\label{2t}

The 2T models provide significantly better fits than the 1T models to
the integrated {\sl ASCA} spectra of all of the groups in our sample
with the exception of NGC 1132. The largest improvements occur for the
groups HCG 62, NGC 5846, and RGH 80 which have among the lowest
average temperatures and the best S/N data in our sample. In Figure
\ref{fig.2t} we display the best-fitting 2T models and residuals for
the same groups appearing in Figure \ref{fig.1t}.  Inspection of
Figure \ref{fig.2t} reveals that with one exception (NGC 5846) the
addition of the second temperature component eliminates most of the
SIS residuals near 1 keV in a manner analogous to the brightest
elliptical galaxies (BF and B99). For NGC 5846 the second temperature
component largely removes the residuals for energies above $\sim 2$
keV. We explore in section \ref{n5846} whether the addition of a third
temperature component can also remove the other SIS residuals for NGC
5846.

Similar to the brightest ellipticals the 2T models also yield values
of $T_{\rm h} \la 2$ keV which indicate that the second temperature
component arises from hot plasma and not the integrated emission from
discrete sources in the central galaxy\footnote{For completeness we
mention that we have also investigated two-component 1T+BREM models
where the temperature of the BREM component is set to a large value
$\sim 10$ keV appropriate for a discrete component as indicated by
some studies (e.g. Matsumoto et al. 1997). For every group where the
2T model outperformed the 1T model we also found that the 2T model
fitted substantially better than the 1T+BREM model analogously to the
brightest ellipticals studied by B99.}. For the systems with the
lowest average temperatures we obtain $\rm EM_h/EM_c \sim 1$ in
agreement with the brightest ellipticals (see B99). This ratio
increases with temperature such that $\rm EM_h/EM_c \sim 4$ for the
hottest groups in our sample (MKW 9 and NGC 4104).

The results listed in Table \ref{tab.fits} for the 2T models apply to
fits that allowed the column densities on each component to vary
separately (except for NGC 6329 -- see section
\ref{comments}). However, the individual columns are generally poorly
constrained for the 2T models. (Note the lower limits on $N_{\rm
H}^{\rm c}$ and upper limits on $N_{\rm H}^{\rm h}$ for many of the
groups). The improvement in the 2T models over the 1T models is not
due to the additional degree of freedom associated with allowing both
$N_{\rm H}^{\rm c}$ and $N_{\rm H}^{\rm h}$ to be free parameters. The
largest improvements are obtained for NGC 4325 and NGC 5846 where
$\chi^2$ decreases by $\sim 5$ when the column densities are untied in
the 2T fits.  Finally, we note that the temperatures and abundance do
not change substantially when the columns are varied separately.

Despite these uncertainties we present individual constraints on
$N_{\rm H}^{\rm c}$ and $N_{\rm H}^{\rm h}$ because for most of the
groups the best-fitting $N_{\rm H}^{\rm c}$ exceeds $N_{\rm H}^{\rm
h}$, the latter of which is typically similar to the Galactic
value. This systematic trend is equivalent to that found for the
brightest elliptical galaxies, though the statistical significance is
much greater for the latter (B99). We comment on the origin of the
excess absorption in section \ref{disc}.

The 2T models give metallicities that are substantially larger than
the 1T models: for the entire sample we obtain $\langle Z\rangle =
0.75\pm 0.24$~$Z_{\sun}$; i.e. the mean is a factor of 2.6 larger than
the 1T value. This systematic increase in $Z$ has also been reported
for the brightest ellipticals by BF and B99 and vividly demonstrates
the sensitivity of the inferred abundance (particularly Fe) to the
assumed temperature structure of the plasma.  This systematic effect,
or bias, is increasingly more important for the systems with lowest
average temperature; e.g., the groups MKW 9 and NGC 4104 which have
the largest average temperatures also have the smallest fractional
increases in $Z$ between the 1T and 2T models. We present a detailed
discussion of this fitting bias in Appendix \ref{bias}.

We mention that even NGC 1132, which shows no improvement in $P$ for
the 2T model, has a metallicity significantly larger than the 1T
value. Moreover, the temperatures, abundance, and other parameters of
the 2T model agree very well qualitatively with those of the other
groups in the sample; i.e. NGC 1132 displays the same fitting bias as
the other groups which strongly suggests that the similarity of its 2T
model parameters to the other groups is not accidental.

When allowing the $\alpha$-process elements to vary separately from Fe
the fits are not improved substantially similar to our results for the
1T models (section \ref{1t}).  However, the same abundances which
improve the 1T fits also improve the 2T models.  Generally, most
improvement occurs when $Z_{\rm O}\rightarrow 0$, and although the
best-fitting $Z_{\rm Si}\rightarrow$ 1.5-2~$Z_{\rm Fe}$ for some
groups, the Si/Fe ratio does not differ within the estimated $1\sigma$
errors from 1. 

The model dependence of the $\alpha$/Fe ratios is best illustrated
with an example. Using 1T models Davis et al. \shortcite{dmm} infer
Si/Fe enhancements for some of their groups which are most significant
for the hottest groups MKW 9 and NGC 4104. When allowing the
$\alpha$-process elements to be separately varied from Fe we obtain a
best-fitting $Z_{\rm Si}/Z_{\rm Fe} = 1.4$ for the 1T model, which
though significantly in excess of 1 is about half the value obtained
by Davis et al.. This 1T model also requires zero abundance for O and
Mg and does not improve the quality of the fit $(P=0.1)$ over the case
where the $\alpha$/Fe ratios are fixed at solar. The 2T model improves
slightly $(P=0.46)$ when varying the $\alpha$/Fe ratios, but $Z_{\rm
Si}/Z_{\rm Fe}$ does not change significantly from 1. The key
improvement arises when letting the O abundance go to zero.

Since the 2T models do not require Si/Fe enhancements for any of the
groups in our sample, and the 2T models are clearly favored over 1T
models, we conclude that there is no evidence for such
enhancements. The inferred peculiar O and Mg abundances inferred by 2T
(and 1T) models suggest that further refinement in the absorption
model (see \ref{disc}), model of the temperature structure, plasma
code, or calibration is required. We mention that if we instead add
another temperature component (i.e. 3T model) with the $\alpha$/Fe
ratios fixed at the solar values typically we are able to obtain
modest improvements in the fits of comparable magnitude to the models
with variable $\alpha$/Fe ratios discussed above (or larger -- see
below in section \ref{n5846}).

\subsubsection{Cooling flow models (CF and CF+1T)}
\label{cf}

From examination of Table \ref{tab.fits} and Figure \ref{fig.cf} one
sees that the multiphase cooling flow models are substantial
improvements over the 1T models and provide fits of comparable quality
to the 2T models for all the groups except for NGC 5846 and RGH 80.
This similarity to the performance of the 2T models is achieved even
though the cooling flow models have one less free parameter due to the
temperatures of the CF and 1T components being tied together (see
section \ref{models}). In fact, because the average temperature of the
CF component is approximately half the upper temperature (e.g. Buote
et al. 1998) the CF+1T models should be similar to 2T models with
$T_{\rm c}\sim 0.5T_{\rm h}$ which applies for the majority of the
groups in our sample.

For HCG 62, NGC 5129, and NGC 6329 only the CF component is required
to produce a fit comparable to the 2T model; i.e. for these groups
(and RGH 80) adding the additional 1T component does not improve the
fits significantly. Note that two of these groups (NGC 5129 and NGC
6329) have the lowest S/N data in our sample.

In Table \ref{tab.fits} the CF model is listed for NGC 5846 even
though the CF+1T model yields a better fit with $P\sim 0.05$. The
primary improvement for the CF+1T model results from allowing the
column density on the hotter component to take a huge value ($N_{\rm
H}^{\rm h}\sim 10^{24}$ cm$^{-2}$) which suppresses the 1T component
for energies below $\sim 2$ keV and thus simulates a higher
temperature component at larger energies. This effect is related to
the SIS and GIS column density discrepancy discussed in section
\ref{issues}. Note if instead we untie the temperatures of the CF and
1T components then a result analogous to the 2T model is
obtained. More complex models for NGC 5846 are considered below in
section \ref{n5846}.

The mass deposition rates $(\dot{M})$ are listed under $\rm EM_c$ for
the cooling flow models in Table \ref{tab.fits}. We obtain typical
$\dot{M}$ values of 5-10 $M_{\sun}$ yr$^{-1}$ for the groups which lie
in between those of giant ellipticals and clusters of galaxies (see
Fabian 1994). Outliers include NGC 2563 whose small $\dot{M}$ is
similar to bright ellipticals (e.g. B99) while the large $\dot{M}$ of
NGC 4325 is similar to clusters.  The contribution of the CF component
to the total emission measure of the CF+1T models is similar to,
though somewhat larger than, the contribution of the colder component
in the 2T models. This is readily apparent from Table \ref{tab.fits}
from comparison of the values of $\rm EM_h$ for the 2T and CF+1T
models.

The column densities obtained for the CF+1T models agree very well
with those obtained for the 2T models. Similar to the 2T models we
also find that the improvement in $\chi^2$ afforded by separate
fitting of $N_{\rm H}^{\rm c}$ and $N_{\rm H}^{\rm h}$ is
insignificant in comparison to that obtained when initially adding the
second temperature component with $N_{\rm H}^{\rm c}=N_{\rm H}^{\rm
h}$. The general trend that $N_{\rm H}^{\rm c}> N_{\rm H}^{\rm h}$
where $N_{\rm H}^{\rm h}$ is similar to the Galactic value is
consistent with cooling flow models of elliptical galaxies (B99) and
clusters of galaxies (e.g. Fabian et al. 1994).

Finally, the metallicities obtained for the cooling flow models
systematically exceed those of the 1T models and are similar to those
obtained for the 2T models: $\langle Z\rangle = 0.65\pm
0.17$~$Z_{\sun}$ for the entire sample. 

\subsubsection{Three-component models for NGC 5846}
\label{n5846}

NGC 5846 represents the group whose {\sl ASCA} spectra are fitted the
poorest by the 2T and cooling flow models: the 2T model gives only a
marginal fit while the cooling flow model is unacceptable (see Table
\ref{tab.fits}). Recall that the addition of the second temperature
component of the 2T model largely removes the residuals for energies
above $\sim 2$ keV but does not significantly affect the SIS residuals
near 1 keV (Figure \ref{fig.2t}). In this section we examine whether
simple extensions of the previous models can eliminate the remaining
SIS residuals for NGC 5846.

It is possible to reduce the SIS residuals to some extent by varying
the $\alpha$/Fe ratios for the 2T model leading to a somewhat better
fit $(P=0.09)$. We find for such a model that the best-fitting
temperatures, column densities, and Fe abundances do not change when
varying the $\alpha$/Fe ratios. In contrast to the other groups (see
section \ref{2t}) we find that all of the $\alpha$-process elements
have best-fitting abundances that exceed Fe, although all the ratios
except for Ne/Fe are consistent with $\alpha$/Fe of solar within the
90\% confidence lower limits.

We can, however, obtain an even better fit when keeping the
$\alpha$/Fe ratios fixed at their solar values if instead we add
another temperature component. In Table \ref{tab.n5846} we give the
best-fitting parameters and their 90\% confidence limits for the 3T
and CF+2T models; the corresponding best-fitting models and residuals
are plotted in Figure \ref{fig.n5846}. For the 3T fits we found that
the column densities on the first and third components were similar
and not well constrained individually, and so we tied them
together. Similarly, for the CF+2T model the column density for the
third temperature component turned out to be small and not well
constrained, so we fixed it to the Galactic value.

These three-component models are formally acceptable and largely
eliminate the SIS residuals near 1 keV similar to the 2T and CF+1T
models of the other groups in the sample. The best-fitting
metallicities are larger than those obtained for the two-component
models but are consistent within the uncertainties. Note in Table
\ref{tab.n5846} that no upper limit is given on $Z$ for the CF+2T
model because the current implementation of the CF model does not
allow $Z>2Z_{\sun}$.

Although the column densities $N_{\rm H}^1$ and $N_{\rm H}^2$ are not
well constrained, both the 3T and CF+2T fits indicate that $N_{\rm
H}^2>N_{\rm H}^1$ in contrast to the two-component models of NGC 5846
and of every other group in the sample. However, similar to the
two-component models the separate fitting of the column densities does
not improve the 3T or CF+2T model fits very much. For example, if the
column densities of each component of the 3T model are tied together
the resulting fit is of essentially the same quality, $P=0.19$,
although some of the fitted parameters change noticeably: best-fitting
$T_1=0.32$ keV, $T_3=2.5$ keV, and $Z=2.7Z_{\sun}$. Nevertheless, it
is possible that the anomalous $N_{\rm H}^2/N_{\rm H}^1$ ratio
reflects complex absorption in NGC 5846, a galaxy with a well-known
and prominent dust lane.

The origin of the third temperature component is unclear. It may arise
in part from discrete sources in the central galaxy, but the
temperature appears to be too low for that to be the only
explanation. An intriguing possibility is that it represents an
extended intragroup gas component discussed by Mulchaey \& Zabludoff
\shortcite{mz}. Since the inferred values of $T_3$ may be too high for
gas in hydrostatic equilibrium in the group potential, but are yet too
small for discrete sources, the third component may well be a
combination of the two effects.

However, as noted above the 2T fits for NGC 5846 can be improved by
varying the $\alpha$/Fe ratios. In fact, we have found that the
parameters of the 3T fits depend on the assumed $\alpha$/Fe
ratios. For example, let us consider the meteoritic solar abundances
of Feldman \shortcite{feld} for which the Fe abundance is a factor of
1.44 larger than the photospheric case. Similar to before we find that
a 3T model fits better ($\chi^2=87.7$/dof=72/$P=0.10$) than the 2T
model, but now the temperature of the second component is $T_2=2$
keV. An even better fit is obtained by replacing the third temperature
component with a BREM component; i.e.  $\chi^2=84.2$/dof=72/$P=0.16$
for the 2T+BREM model. For each of these (meteoritic) 3T and 2T+BREM
models the fitted parameters for the lower temperature components are
more similar to those of the other groups (Table \ref{tab.fits}) and
the brightest ellipticals (B99). For example, for the 2T+BREM model we
obtain best-fitting parameters $N_{\rm H}^1=1.9\times 10^{21}$
cm$^{-2}$, $N_{\rm H}^2=0$, $T_1=0.62$ keV, $T_2=1.66$ keV, $T_{\rm
BREM}= 3.98$ keV, $Z=2.79Z_{\sun}$, and $EM_1:EM_2:EM_{\rm BREM} =
7.4:2.0:0.17$. Hence, using meteoritic abundances the parameters of
the 2T+BREM model for NGC 5846 are analogous to those of the brightest
elliptical galaxies (B99), and thus the third spectral component
(i.e. BREM) is suggestive of emission from discrete sources in this
case. 

Finally, we mention that the other groups do not show as much
improvement as NGC 5846 for 3T and CF+2T models because the SIS
residuals near 1 keV are already mostly eliminated by the 2T and CF+1T
models (sections \ref{2t} and \ref{cf}). The marginal fits for a few
of the other groups are the result of other residuals (e.g. for HCG 51
at high energies) that are probably the result of errors in the
background subtraction or calibration, though errors in the
temperature model and plasma code also may contribute.

\subsubsection{Comments on individual groups}
\label{comments}

The spectral fitting of some of the groups require additional
comment.

\medskip

\noindent {\bf HCG 51:} If we exclude the energies above 3 keV the
fits are improved significantly; e.g. $P=0.19$ for the 2T model.

\noindent {\bf MKW 9:} The SIS0 and SIS1 data are very inconsistent
below 0.7 keV such that models always lie significantly below the SIS0
and above the SIS1 data. As a result we excluded SIS0 and SIS1 data
below 0.7 keV.

\noindent {\bf NGC 533:} Trinchieri, Fabbiano, \& Kim \shortcite{tfk}
fitted a 2T model to their {\sl ROSAT} PSPC spectra accumulated within
a radius of $6\arcmin$. Although they tied together the column
densities on both components and fixed the Fe abundance at solar, they
also found that the 2T model fitted better than an isothermal model
and obtained values for $T_{\rm c}$ and $T_{\rm h}$ in good agreement
with ours.

\noindent {\bf NGC 1132:} We note that the best-fitting 1T abundance
of $0.58Z_{\sun}$ obtained by Mulchaey \& Zabludoff \shortcite{mz1132}
using the Raymond-Smith code is about twice the value listed in Table
\ref{tab.fits}. We reproduce their result when using the Raymond-Smith
code.

\noindent {\bf NGC 5846:} Our 1T and CF models are consistent with
those published by BF, but the values of $T_{\rm h}$ and $Z$ for our
2T model are larger than those obtained by BF. Roughly equal
contributors to these differences are that BF (1) set $N_{\rm H}^{\rm
c}=N_{\rm H}^{\rm h}$ and (2) did not include the GIS data in the
fits.

\noindent {\bf NGC 6329:} Since $N_{\rm H}^{\rm c}$ is unconstrained
within the estimated 90\% uncertainties we tied together the column
densities in the 2T fits.

\section{Discussion}
\label{disc}

\subsection{Comparison with previous work}
\label{prev}

The results we have obtained for the metal abundances in the hot gas
of poor groups of galaxies differ markedly from those obtained by
previous X-ray studies. In this section we discuss more generally how
our findings compare with previous studies. As a consequence of this
discussion we present a synthesis of the properties of the hot gas in
galaxy groups below in section \ref{prop}.

We have focused our analysis of poor galaxy groups on the integrated
{\sl ASCA} spectra accumulated within apertures of $\sim
3\arcmin$-$5\arcmin$ radius (i.e. $\sim 50$ - 200 kpc). For the groups
in our sample these apertures typically enclose $\sim 50$ per cent of
the total X-ray emission and, perhaps more importantly, $\ga 90$ per
cent of the flux within the projected aperture radius, $R_{\rm ap}$,
originates from within the three-dimensional radius $r=R_{\rm ap}$;
i.e. essentially our analysis is that of the integrated spectra
accumulated within the three-dimensional radius $r\sim 50$ - 200 kpc
of each group.

These region sizes were chosen as a compromise between S/N and a
desire to enclose a large fraction of the group emission.  Since the
X-ray surface brightness decreases rapidly with increasing radius the
ability to constrain an arbitrary spectral model diminishes
accordingly. Some authors have avoided the bright central region
because models more complex than a single isothermal component are
often required for, in particular, galaxy clusters, whereas the lower
S/N spectra at larger radii usually require only a single isothermal
component. We prefer to focus on the regions which provide the
strongest model constraints even though their spectra are expected to
be more complex than at larger radii (because of, e.g., cooling
flows).

Previous X-ray studies of groups of galaxies that fitted a single 1T
model (e.g. Fukazawa et al. 1996, 1998; Davis et al. 1999) to the
integrated {\sl ASCA} spectra over a large region in every case
obtained very sub-solar Fe abundances and for many cases obtained
Si/Fe ratios in excess of solar. However, we have shown that the {\sl
ASCA} spectra accumulated within apertures of $\sim
3\arcmin$-$5\arcmin$ radius (i.e. $\sim 50$ - 200 kpc) strongly favor
models consisting of at least two temperature components for 11 of the
12 groups in our sample. Moreover, the Fe abundances determined for
the multiphase models (2T or cooling flow) are nearly solar and the
$\alpha$/Fe ratios are consistent with the solar values. As a result
for studies like Fukazawa et al. \shortcite{fuk96}, who analyzed the
accumulated {\sl ASCA} spectra of HCG 62 within an aperture of
$5\arcmin$ radius, we conclude that their very sub-solar Fe abundance
and Si/Fe ratio in excess of solar are clearly an artifact of forcing
isothermal models to fit the multitemperature {\sl ASCA} spectra.
This fitting bias is examined in detail in Appendix \ref{bias}.  (The
same holds for {\sl ROSAT} studies such as Ponman et al. 1996 and
Mulchaey \& Zabludoff 1998 who found very sub-solar metallicities
within radii of $\sim 200$ kpc of several poor groups.)

The metal abundances obtained by both Fukazawa et
al. \shortcite{fuk98} and Davis et al. \shortcite{dmm} also agree
quite well with the values we obtained using isothermal models (Table
\ref{tab.fits}), and thus are consistent with this fitting
bias. However, it is possible that a strong metallicity gradient may
in part explain the very sub-solar Fe abundances obtained by Fukazawa
et al. and Davis et al.  who analyzed the accumulated {\sl ASCA}
spectra within radii that are typically factors of 5-10 larger than
ours. This explanation is more plausible for Fukazawa et
al. \shortcite{fuk98} since they include emission only within an
annulus of $0.1$-$0.4$ $h_{50}^{-1}$ Mpc for each group.  For example,
for HCG 62 Fukazawa et al. obtain $Z_{\rm Fe}=0.15\pm 0.03Z_{\sun}$
within an annulus 70-286 $h^{-1}_{70}$ kpc as opposed to our value of,
e.g., $0.99_{-0.38}^{+0.94}Z_{\sun}$ for the 2T model within a radius
of 55 $h^{-1}_{70}$ kpc. In order for these values to match there must
be a steep metallicity gradient. The required gradient seems to be
implausibly steep since (1) our multitemperature spectral models (2T
and CF+1T) did not require different metallicities on the different
spectral components for any group in our sample, and (2) below in
section \ref{abun} we show that the $Z$-$R_{\rm ap}$ correlation
argues against a substantial decrease in metallicity out to radii of
at least $\sim 200$ kpc in galaxy groups.

It is even more doubtful that abundance gradients can give a full
accounting of our differences with Davis et al. because our $\sim 100$
kpc apertures enclose typically $\sim 50$ per cent of the emission
within the apertures used by Davis et al.. If we again consider the
case of HGC 62, Davis et al. obtained $Z=0.21^{+0.03}_{-0.04}Z_{\sun}$
within a radius of $23\arcmin$ which is only slightly larger than
Fukazawa et al.'s value of $Z_{\rm Fe}=0.15\pm 0.03Z_{\sun}$ obtained
from a region that excludes the center but includes the extended
emission present in the apertures of Davis et al.. If abundance
gradients are to account for our differences, then Davis et al. should
have obtained a weighted average of our metallicity
($0.99_{-0.38}^{+0.94}Z_{\sun}$) and that of Fukazawa et al.. This
would amount to $Z\sim 0.6Z_{\sun}$ since our region and Fukazawa et
al.'s region have approximately the same fluxes. This value is well in
excess of that obtained by Davis et al.. Even in the extreme case that
$Z=0$ outside of our aperture, we would have expected Davis et al. to
measure $Z\sim 0.5Z_{\sun}$.

Therefore, even if $Z_{\rm Fe}\ll 1Z_{\sun}$ at radii much larger than
our apertures, the ``Fe bias'' is very likely an important contributor
to the very sub-solar Fe abundances measured by Fukazawa et
al. \shortcite{fuk98} and especially Davis et al. \shortcite{dmm}.

These same arguments also apply for the Si/Fe enhancements found for
some groups by Fukazawa et al. \shortcite{fuk98} and Davis et
al. \shortcite{dmm}; i.e. we find that Si/Fe enhancements can be
explained by a bias (see section \ref{si}) resulting from fitting an
isothermal model to an intrinsically multitemperature spectrum,
however it is possible that abundance gradients partially could
account for the fact that Fukazawa et al. and Davis et al. measure
Si/Fe enhancements in some groups whereas our multitemperature models
do not require Si/Fe ratios different from solar.

It should be remarked that these previous studies have used the
photospheric solar abundances whereas, as emphasized by Ishimaru \&
Arimoto \shortcite{im}, the meteoritic solar abundances are more
appropriate. (Recall that we have used the photospheric solar
abundances to facilitate comparison with these previous studies --
section \ref{models}.) For comparison we have also performed the
spectral fits for the groups with the highest S/N in our sample
(i.e. those in Figures \ref{fig.1t} - \ref{fig.cf}) using the
meteoritic solar abundances of Feldman \shortcite{feld}. As would be
expected we obtain fits of comparable quality to before, but now the
values of the Fe abundances are raised by a factor of $\sim 1.44$. As
before we also find that the 2T models do not generally require a
Si/Fe ratio different from (now meteoritic) solar within the 90 per
cent confidence limits. However, the best-fitting values of the Si/Fe
abundance ratio now tend to be 10-20 per cent {\it below}
solar. Hence, within our $\sim 100$ kpc apertures we find no evidence
for Si/Fe enhancements, and when meteoritic abundances are used if
anything we detect weak evidence for Si/Fe decrements.

For several groups in our sample the temperature profiles measured
from spatially resolved spectral analysis of the {\sl ROSAT} PSPC
(e.g. Trinchieri et al. 1997; Mulchaey \& Zabludoff 1998) and {\sl
ASCA} (Finoguenov et al. 1999; Finoguenov \& Ponman 1999) data provide
corroborating evidence for multi-temperature hot gas in groups of
galaxies. The temperature profiles are qualitatively similar for each
group: starting from a minimum at the center the temperature rises
sharply until reaching a maximum at radii of a few arcminutes and then
falls gently at larger radii. These qualitative features are entirely
analogous to those of the brightest elliptical galaxies; see, e.g.,
the compilation by Brighenti \& Mathews \shortcite{bmtprof}. Given the
similarity of the X-ray spectra of poor groups and bright elliptical
galaxies it should be expected that the very sub-solar Fe abundances
of elliptical galaxies found by most previous X-ray studies also arise
from the fitting bias associated with the assumption of isothermal gas
(BF; B99).

The temperature maxima occur almost precisely at the edges of the
apertures we use to extract the {\sl ASCA} spectra (Table
\ref{tab.obs}); i.e. our apertures include the regions enclosing the
positive temperature gradients in the groups (analogously to our
previous studies of ellipticals). Plausible physical explanations of
these rising temperature profiles are, e.g., a single-phase medium
resting in a hierarchical gravitational potential, a two-phase medium,
or a multiphase cooling flow; below in section \ref{2tvscf} we discuss
these scenarios and how they may be distinguished with future
observations. 

B99 showed that the (positive) {\sl ROSAT} temperature gradients of
the bright ellipticals NGC 1399, NGC 4472, and NGC 5044 are
inconsistent with the isothermal models obtained from fitting the
integrated {\sl ASCA} spectra accumulated within apertures of $\sim
5\arcmin$ radius but are very consistent with simple two-temperature
and cooling flow models. (Note the X-ray emission of NGC 5044 is
better described as that originating from a group -- see section
\ref{abun}.)  Since the {\sl ROSAT} temperature profiles and {\sl
ASCA} multitemperature models for the groups are so similar to those
of the brightest ellipticals, we refer the interested reader to
section 5 of B99 for a detailed demonstration of their
consistency. This ``equivalence'' of a 2T model and temperature
gradient was also shown by Trinchieri et al. \shortcite{tfk} for the
NGC 533 group using {\sl ROSAT}. As mentioned in section
\ref{comments} our 2T model for NGC 533 (Table \ref{tab.fits}) agrees
well with that of Trinchieri et al..

For our present discussion we emphasize that the parameters obtained
from fitting two-temperature and cooling flow models to the integrated
spectra within a single aperture are self-consistent in the sense that
the parameters accurately represent the values of the radially varying
parameters within that aperture. We demonstrated this consistency via
simulated {\sl ASCA} SIS observations of the brightest ellipticals in
section 5.3 of B99. Specifically, for each elliptical B99 constructed
a series of 1T models from the {\sl ROSAT} data as a function of
radius within the radius of the {\sl ASCA} aperture. B99 summed these
radially varying models to form a composite emission model within the
{\sl ASCA} aperture and then simulated an {\sl ASCA} SIS observation
with the same exposure as the real observation.

B99 demonstrated that the values of $T_{\rm c}$ and $T_{\rm h}$
deduced from fitting the simulated SIS data with two-temperature
models agreed very well with respectively the lowest and highest
temperatures of the input radially varying {\sl ROSAT} models. More
importantly for our present discussion, the metallicity obtained from
the 2T fit reflected an average value of the metallicities of the
input {\sl ROSAT} models. This same level of agreement was also found
for the cooling flow models. This exercise shows that the
metallicities obtained by two-temperature and cooling flow models
fitted to the accumulated {\sl ASCA} spectra within radii enclosing
the regions of positive temperature gradient reflects an average of
the metallicities over the radii within the aperture. (The consistency
of the {\sl ROSAT} and {\sl ASCA} models fitted to the real data is
discussed in section 5.2 of B99.)

As a result, the metallicities of our 2T and cooling flow models
obtained for poor groups of galaxies are inconsistent with those
obtained by Finoguenov et al. \shortcite{fino} and Finoguenov \&
Ponman \shortcite{fp} within the radii we have examined (Table
\ref{tab.obs}). These authors performed a spatially resolved spectral
analysis of {\sl ASCA} and {\sl ROSAT} data for the groups HCG 51, HGC
62, and NGC 5846 (also NGC 5044) and obtained very sub-solar Fe
abundances and Si/Fe ratio enhancements that agree very well with the
previous studies cited earlier in this section that simply fitted
isothermal models to the integrated {\sl ASCA} spectra of these
groups; recall that the abundances obtained from those studies suffer
from the fitting bias described in Appendix \ref{bias}. Within the
radii of our apertures the Fe abundances obtained by Finoguenov et
al. and Finoguenov \& Ponman are $Z_{\rm Fe}<0.4Z_{\sun}$ for HCG 51,
$Z_{\rm Fe}<0.4Z_{\sun}$ for HCG 62, and $Z_{\rm Fe}<0.3Z_{\sun}$ for
NGC 5846: These very sub-solar Fe abundances cannot be reconciled with
the metallicities we have obtained for these groups using
multitemperature models (Tables \ref{tab.fits} and \ref{tab.n5846}).

Since our two-temperature and cooling flow models accurately reflect
the average metallicity within the apertures enclosing the region of
positive temperature gradient, and since the two-temperature and
cooling flow models (whose differential emission measures have very
different shapes) generally give similar results for the abundances,
we conclude that the metallicities we have derived are robust and thus
$Z>0.5Z_{\sun}$ (90 per cent confidence) within radii of $\sim 100$
kpc when averaged over the groups in our sample.

In contrast, we are unable to assess the reliability of the {\sl ASCA}
deconvolution analysis of Finoguenov et al. \shortcite{fino} and
Finoguenov \& Ponman \shortcite{fp} since they do not demonstrate the
validity of their procedure using simulated data. There are several
assumptions made in their procedure (e.g. no correlation between
spatial and spectral properties -- see equation 3 of the Appendix in
Finoguenov et al., the regularization assumptions in section 2 of
Finoguenov \& Ponman, and the assumption that the gas is single-phase
everywhere) which could potentially impact their results. As a
substitute to simulation, these authors present analogous results
using {\sl ROSAT} data which (in Finoguenov \& Ponman) are obtained by
using an independent method.  Although the abundances obtained by
Finoguenov et al and Finoguenov \& Ponman for the {\sl ROSAT} data of
HGC 62, NGC 5044, and NGC 5846 are factors of 1.5-2 times larger than
those obtained using the {\sl ASCA} data, these systematic differences
are dismissed because they are marginally consistent within the 90 per
cent confidence limits with the {\sl ASCA} results. (Note no {\sl
ROSAT} results are presented for HCG 51.)

The systematic differences of the {\sl ROSAT} and {\sl ASCA} Fe
abundances obtained by Finoguenov et al and Finoguenov \& Ponman
indicate that systematic errors are significant either for their {\sl
ASCA} results, their {\sl ROSAT} results, or both. There is ample
reason to question the reliability of their method to deconvolve {\sl
ASCA} data because the results which have been obtained for groups and
galaxy clusters are very sensitive to the deconvolution method used
(e.g. Markevitch et al. 1998; Kikuchi et al. 1999; for a review of
this subject see Irwin, Bregman, and Evrard 1999). Moreover, the PSF
should not be very important for the ROSAT analysis but it is crucial
for the {\sl ASCA} analysis; recall that the radii under consideration
are $3\arcmin$-$5\arcmin$, and the half-power diameter of the {\sl
ASCA} XRT is $\sim 3\arcmin$, so the annuli analyzed by Finoguenov et
al and Finoguenov \& Ponman within our apertures are considerably
smaller than the {\sl ASCA} PSF.

Hence, because Finoguenov et al and Finoguenov \& Ponman do not
demonstrate the reliability of their {\sl ASCA} deconvolution via
simulation, and they obtain systematic differences in metallicities
between the {\sl ROSAT} and {\sl ASCA} data (with the {\sl ROSAT}
metallicities being systematically larger and in better agreement with
our single-aperture analysis) we suspect that the abundances
determined by these authors within our apertures using {\sl ASCA} data
are inconsistent with our results owing to systematic errors in their
deconvolution procedure.

\subsection{Properties of the hot gas in groups of galaxies}
\label{prop}

Therefore, very similar to the brightest ellipticals (B99 and BF) the
X-ray spectral properties in the central regions ($\sim 100$ kpc) of
groups of galaxies can be summarized as follows:

\begin{enumerate}
\item The hot gas within the {\sl ASCA} apertures ($\sim 100$ kpc)
consists of at least two temperature components.

\item The average temperature is maximum near the edge of the {\sl
ASCA} apertures and declines monotonically to a minimum value at the
center as indicated by both {\sl ROSAT} and {\sl ASCA} studies.

\item The Fe abundances are $\sim 0.7Z_{\sun}$ and the $\alpha$/Fe
ratios are consistent with their (photospheric) solar values. These
abundances represent values averaged over the {\sl ASCA} apertures and
thus may reflect abundance gradients within. Though abundance
gradients within the {\sl ASCA} apertures are not required by our
multicomponent spectral models\footnote{That is, for each group the
abundances of each spectral component are generally consistent with
each other.}, the $Z$-$R$ correlation discussed in section \ref{abun}
is suggestive of $Z$ declining with increasing radius.

\item The two-component {\sl ASCA} models suggest absorption in excess
of the Galactic value on the colder temperature component while the
column density on the hotter component is generally consistent with
Galactic. Hence, the excess absorption is concentrated at the center
of the group.

\end{enumerate}  

Although the measured excess absorption for the groups (sections
\ref{2t} and \ref{cf}) is not as significant as for the brightest
ellipticals (B99) or for clusters of galaxies (e.g. Fabian et
al. 1994), the impression is generally the same; i.e. $N_{\rm H}^{\rm
c} > N_{\rm H}^{\rm h}$ and $N_{\rm H}^{\rm h}$ is similar to the
Galactic value. The large values of $N_{\rm H}^{\rm c}$ for
ellipticals and clusters imply large amounts of intrinsic absorbing
material that generally exceed the amount of cold gas inferred from HI
or CO observations of ellipticals (e.g. Bregman, Hogg, \& Roberts
1992) and clusters (e.g. O'Dea et al. 1994). Although excess
absorption is consistent with the standard multiphase cooling flow
models with mass drop-out (e.g. Fabian 1994), this discrepancy between
X-ray observations and observations at other wavelengths poses a
serious challenge to our understanding of the hot gas in these
systems. 

Finally, we note that our analysis does not probe the outer regions
(i.e. $r\ga 200$ kpc) of these groups, and thus it is possible that
the hot gas in the outer regions of groups has very sub-solar Fe
abundances and Si/Fe ratios above (photospheric) solar as suggested in
particular by Fukazawa et al. \shortcite{fuk98} and Finoguenov \&
Ponman \shortcite{fp}. However, we regard their evidence as tentative
for the reasons discussed in section \ref{prev}; i.e. isothermal
modeling by Fukazawa et al. and systematics in the {\sl ASCA}
deconvolution procedure by Finoguenov \& Ponman.

\subsection{Two-temperature vs. cooling flow models}
\label{2tvscf}

Similar to the brightest elliptical galaxies the {\sl ASCA} spectra of
galaxy groups accumulated within radii of $\sim 3\arcmin$-$5\arcmin$
do not distinguish clearly between two-temperature and simple
multiphase cooling flow models; i.e. whether the hot gas emits at only
two temperatures or from a range of temperatures cannot be determined
presently.  This inability of the {\sl ASCA} data to discriminate
between 2T and cooling flow models is expected even if the cooling
flow models are correct, because the 2T models have an additional free
parameter and can approximate very well a cooling flow spectrum
throughout the {\sl ASCA} bandpass (Buote, Canizares, \& Fabian 1999).
In this section we discuss possible physical scenarios for the
two-temperature models and at the end briefly mention the prospects
for distinguishing 2T models from cooling flows with future X-ray
observations.

The most widely explored two-phase model emphasizes the role of a
``hierarchical potential'' structure in the hot gas of ellipticals,
groups, and clusters. This model ascribes the central enhancements in
X-ray surface brightness and central depressions in X-ray temperature
to the fact that the central gas sits in the shallower potential well
associated with the halo of the cD galaxy while the more extended
hotter gas is associated with the gravitational potential of the
surrounding group or cluster. In particular, for the clusters of
galaxies Hydra-A \cite{ikebe97} and A1795 \cite{xu} it has been shown
that the two-phase models fit the {\sl ASCA} data as well as
multiphase cooling flow models. (Typically cooling flow models of
clusters also incorporate the influence of the central galaxy on the
gravitational potential; e.g. Thomas, Fabian, \& Nulsen 1987; White \&
Fabian 1995)

This ``hierarchical potential'' model has also been invoked to explain
the {\sl ASCA} data of the Fornax cluster which houses the bright
elliptical NGC 1399 at its center \cite{ikebe96}. However, this
two-phase model cannot explain the multi-temperature structure implied
by the 2T models obtained by BF and B99 for the {\sl ASCA} data within
a radius of $\sim 5\arcmin$. Firstly, Ikebe et al. assume both
components are isothermal. Secondly, the extended component
corresponding to the group halo contributes only a small faction of
the emission within a $\sim 5\arcmin$ radius whereas the colder and
hotter temperature components obtained by the 2T models of BF and B99
essentially contribute equally to the emissivity. This also applies to
the groups in our sample for which Mulchaey \& Zabludoff
\shortcite{mz} have found similar two-component structure in the
radial distribution of the X-ray emission\footnote{It is possible that
the third temperature component for the 3T and CF+2T models of NGC
5846 may arise in part from an extended group component (section
\ref{n5846}).}.

A two-phase model which can explain the temperature structures implied
by the 2T fits to the {\sl ASCA} spectra of the brightest ellipticals
and groups is described by Brighenti \& Mathews (1998; 1999a).
Brighenti \& Mathews \shortcite{bminflow} assume that primordial gas
at large radii is shock-heated to the virial temperature of the
surrounding group potential as the gas flows in to the center during
secondary infall. This hotter gas phase is similar to the extended
component invoked in the ``hierarchical potential'' models above.  In
addition, Brighenti \& Mathews account for the mass ejected from stars
in the central galaxy during the course of normal stellar
evolution. This process continually ejects gas preferentially at the
center of the system with a temperature appropriate to the
gravitational potential of the central galaxy; i.e. the ejected gas
from stellar mass-loss is cooler than the ambient shock-heated
primordial gas.

Brighenti \& Mathews \shortcite{bminflow} have shown that this
two-phase model can accurately reproduce the {\sl ROSAT} temperature
and surface brightness profiles of the bright elliptical NGC
4472. Since we demonstrated in B99 that the {\sl ROSAT} temperature
gradients for four of the brightest ellipticals (including NGC 4472)
are very consistent with the 2T models obtained from fitting the {\sl
ASCA} spectra, we conclude that the scenario described by Brighenti \&
Mathews can also explain the 2T models obtained from fitting the {\sl
ASCA} spectra of ellipticals and groups (BF; B99; this paper).

The model of Brighenti \& Mathews is a {\it bona fide} two-phase
system whereas the above ``hierarchical potential'' model is truly
single-phase (but not necessarily isothermal) in the sense that a
single value of the density and temperature exists at each
radius. Although it has no effect on their calculations, it is assumed
by Brighenti \& Mathews that the stellar ejecta and ambient hot gas
mix instantaneously so that a single-phase medium results. However,
near the center where the injection of stellar material is most
important this may not be the case, and the two phases may co-exist.

In contrast, a continuum of phases is predicted to exist in the cores
of ellipticals, groups, and clusters in the standard multiphase
cooling flow scenario (Fabian, Nulsen, \& Canizares 1984; Nulsen 1986,
1998; Thomas et al. 1987; White \& Sarazin 1987; see Fabian 1994 for a
review). This scenario assumes that in the regions of highest density
where the cooling time of the ambient gas is sufficiently short
(usually less than the assumed cluster age) gas cools rapidly below
X-ray temperatures and drops out of the flow typically with a mass
deposition profile $\dot{M}(<r)\sim r$.  The fate of the cooling gas
remains the most serious problem for this scenario (e.g. Fabian 1994),
though the cooling gas could provide a natural explanation of the
excess absorption obtained from fitting multi-temperature spectral
models to the {\sl ASCA} spectra of ellipticals, groups, and
clusters. Although excess absorption is not an obvious prediction of
the other models described above, Pellegrini \& Ciotti \shortcite{pc}
argue that single-phase cooling flows with partial winds can produce
cold gas that would give rise to excess absorption.

It is probably inevitable that aspects of both the two-phase models
and multiphase cooling flow scenario operate to some extent in
ellipticals, groups, and clusters: i.e. gas cooling out of the hot
phase, cool gas ejected from stars in the cD, and hotter gas due to
primordial gas shock-heated to the virial temperature of the
surrounding group or cluster.  Data from future X-ray missions will be
able to distinguish between 2T and cooling flow models from spectral
fitting within a large spatial aperture \cite{b98}. Moreover, the
vastly improved spatial resolution of the {\sl Chandra} and {\sl XMM}
satellites over previous missions will allow the location of excess
absorbing material and the spatial distribution of distinct gas phases
(if present) to be mapped with unprecedented accuracy.

\subsection{Implications of nearly solar abundances}
\label{abun}

\begin{figure}
\centerline{\hspace{0cm}\psfig{figure=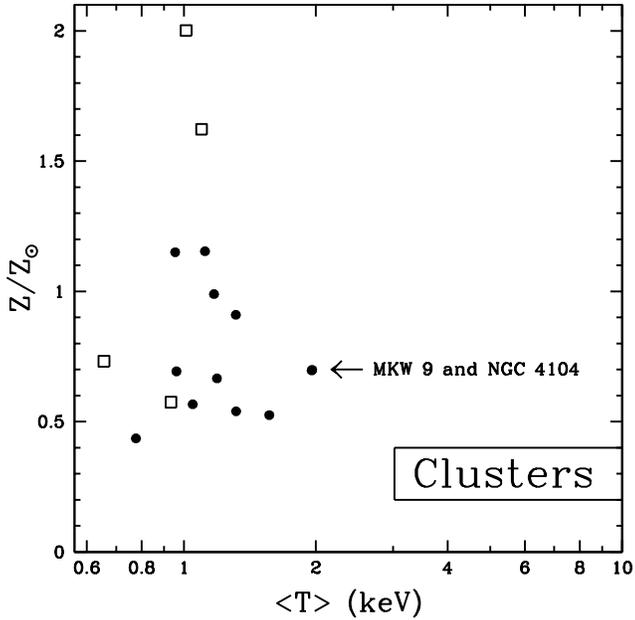,width=0.49\textwidth,angle=0}}
\caption{\label{fig.abun} The metallicities and the
emission-measure-weighted temperatures for ellipticals and groups
obtained from two-temperature spectral models. The filled circles
represent the groups from this paper while the open squares refer to
the bright ellipticals NGC 1399, NGC 4472, NGC 4636, and NGC 5044
analyzed by B99. The region denoted ``Clusters'' indicates the typical
area populated by rich clusters. See section \ref{abun} for details.}

\end{figure}

\begin{figure*}
\parbox{0.49\textwidth}{
\centerline{\psfig{figure=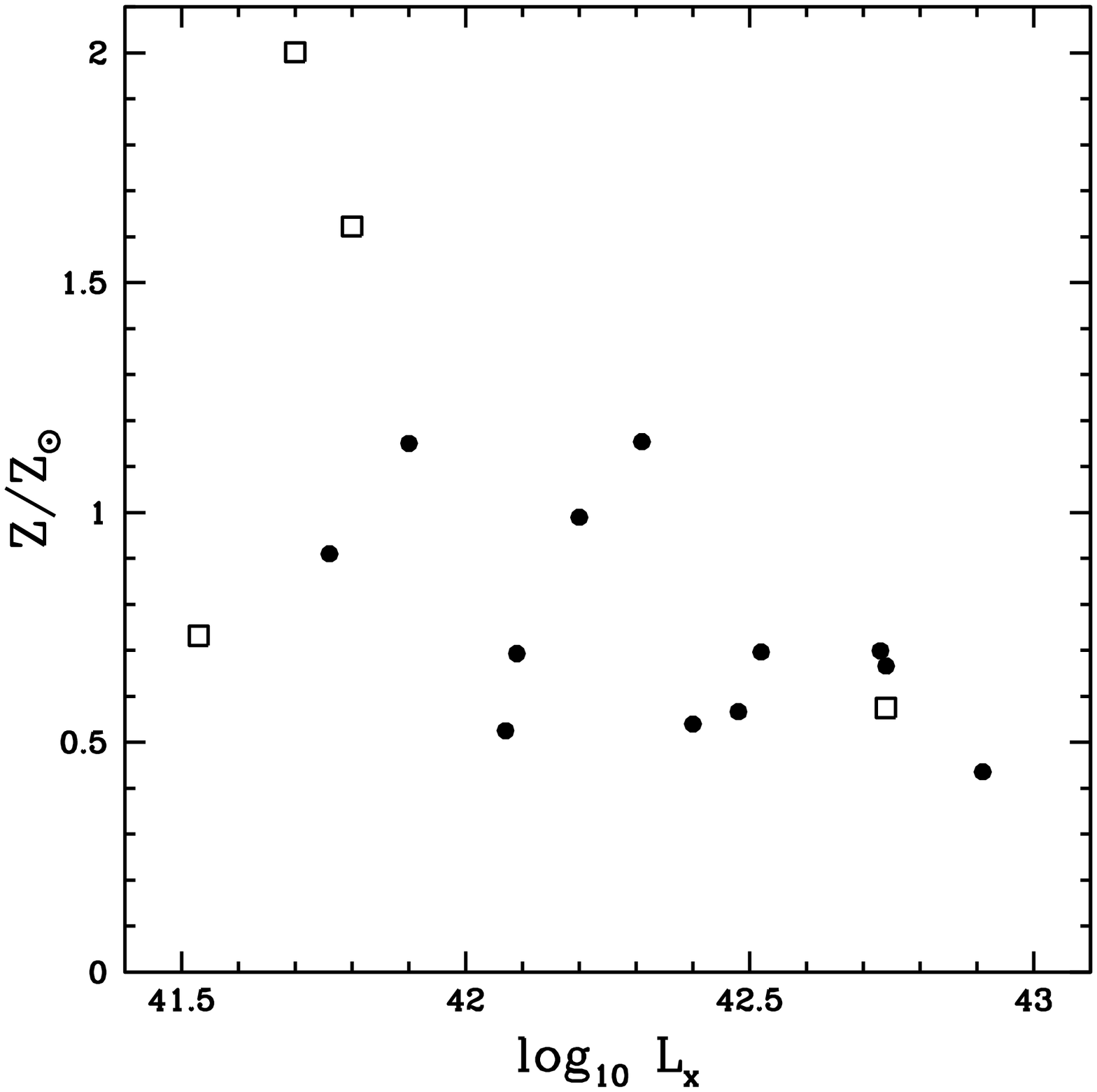,angle=0,height=0.33\textheight}}
}
\parbox{0.49\textwidth}{
\centerline{\psfig{figure=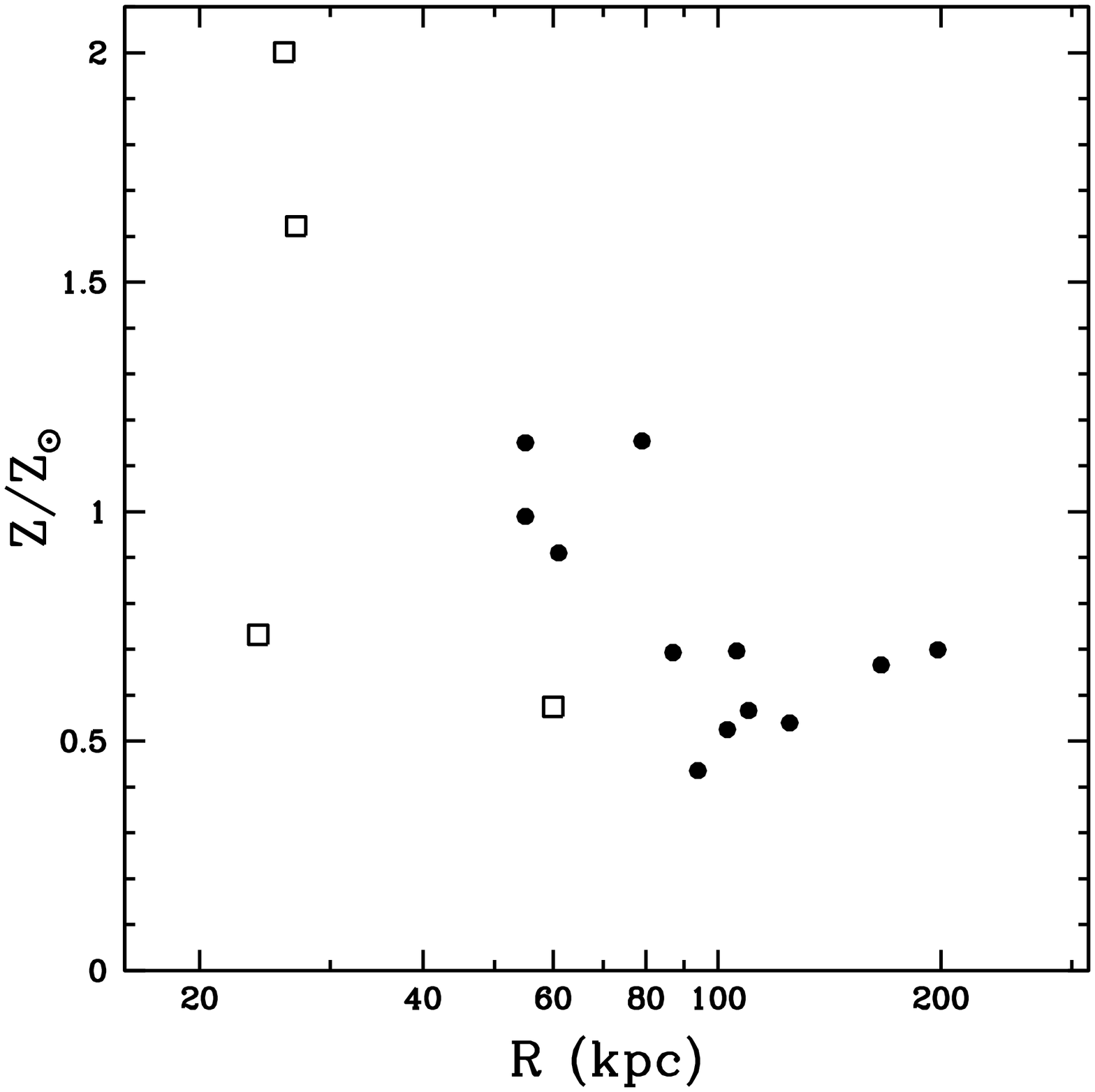,angle=0,height=0.33\textheight}}
}
\caption{\label{fig.abungroup} As Figure \ref{fig.abun} except here we
plot the metallicity versus (left) the (unabsorbed) luminosity in the
0.5-10 keV band (see Table \ref{tab.prop}) and (right) the size of the
aperture used for spectral analysis.}
\end{figure*}

In Figures 5 and 6 of Renzini \shortcite{renzini} it is shown that the
Fe abundances of groups and clusters determined from X-ray
observations are much more strongly correlated with temperature than
with the luminosity $L_B$. For systems above $T\sim 2$ keV the Fe
abundance is approximately constant at $\sim 0.3Z_{\sun}$. Below
$T\sim 2$ keV there is considerable scatter, but most of the systems
have $Z_{\rm Fe}\la 0.2Z_{\sun}$. The Fe abundances of groups quoted
by Renzini \shortcite{renzini} were obtained from published isothermal
spectral models of the {\sl ROSAT} or {\sl ASCA} data in the
literature and therefore suffer from the fitting bias discussed in
Appendix \ref{bias}.

We can achieve a more consistent comparison using the 1T models
obtained for the groups in our sample. The average metallicity and
$1\sigma$ range for the groups in our sample deduced from 1T models
(section \ref{1t}), $\langle Z\rangle = 0.29\pm 0.12$~$Z_{\sun}$, is
very consistent with those obtained for galaxy clusters.  Similar to
Figure 6 of Renzini \shortcite{renzini}, $Z$ tends to decrease with
decreasing temperature for our 1T models. This trend evidently
continues below $\sim 1$ keV since, e.g., BF obtained $\langle
Z\rangle = 0.19\pm 0.12$~$Z_{\sun}$ from 1T fits of a sample of 20
bright ellipticals.

Many have speculated that since $Z$ correlates with $T$ below $\sim 2$
keV, whereas $Z$ is approximately constant at higher temperatures,
there is likely a systematic error affecting the measurement of the Fe
abundances. Probably the most frequent suggestion (e.g. Arimoto et
al. 1997; Renzini 1997) is that the systematic trend is due to errors
in the Fe L lines of the plasma codes, but previously we have shown
that existing Fe L errors are not serious for temperature and
abundance determination of normal ellipticals which should be the most
affected (BF; B99). Our previous analyses of ellipticals and our
current analysis of groups instead demonstrate that the key systematic
error in the determination of $Z$ results from assuming the gas to be
isothermal in these systems.

In Figure \ref{fig.abun} we plot $Z$ versus emission-measure-weighted
temperature obtained from the 2T models for the groups in our sample.
For comparison we also show in Figure \ref{fig.abun} the region
containing most of the cluster $Z$ measurements as compiled by Renzini
\shortcite{renzini}. We note that these cluster metallicities were
obtained from {\sl GINGA} and {\sl EXOSAT} data and thus include
emission over a large portion of the clusters whereas our results for
groups are obtained within apertures of $\sim 100$ kpc which typically
contain $\sim 50$ per cent of the total X-ray emission.

The trend is now quite different from Renzini's Figure 6 since below
$\sim 2$ keV we find that $Z$ {\sl increases} to $\sim 0.7Z_{\sun}$
for groups. Unlike the 1T models there is no obvious correlation of
$Z$ with $\langle T\rangle$ for the 2T models; i.e. removing the ``Fe
bias'' with the 2T models also largely removes the dependence of $Z$
on temperature for groups.

The 2T models of ellipticals (BF) suggest that $Z$ may continue to
increase for lower temperatures, though the scatter is considerable,
$\langle Z\rangle = 0.9\pm 0.7$~$Z_{\sun}$. Let us instead focus on
the ellipticals NGC 1399, NGC 4472, NGC 4636, and NGC 5044 studied by
B99 since they have the best available constraints on the spectral
models. Moreover, for NGC 1399, NGC 4472, and NGC 4636 B99 included an
emission component that accounts for emission from discrete sources as
expected in these systems. The results for these galaxies are plotted
as open squares in Figure \ref{fig.abun}. These objects overlap in
temperature with the groups and add extra scatter in $Z$.

Since $Z$ and $\langle T\rangle$ do not exhibit a strong correlation
for the groups and ellipticals we investigate correlations of $Z$ with
other physical quantities. In Figure \ref{fig.abungroup} we plot $Z$
versus $L_{\rm x}$ in the 0.5-10 keV band. There is a significant
correlation in the sense that more luminous systems have lower
$Z$. This trend for groups appears to connect smoothly with clusters
$(Z\sim 0.3Z_{\sun})$ at high $L_{\rm x}$.  We mention that similar
correlations exist if instead of $L_{\rm x}$ we plot $Z$ against the
emission measure of the cooler component (i.e. $\rm EM_c \times 4\pi
D^2$) or $\dot{M}$ for the cooling flow models; i.e. the trend of
decreasing $Z$ with increasing $L_{\rm x}$ (i.e. mass) can also be
expressed as $Z$ decreasing either with increasing luminosity of the
cooler component of the 2T models or with increasing mass deposition
rate for the cooling flow models.

It is clear from Figure \ref{fig.abungroup} that with respect to X-ray
emission the ``elliptical'' NGC 5044 $(\log_{10}L_{\rm x} = 42.74)$ is
more accurately described as a bright group which is also indicated by
its large $\dot{M}$ obtained from a cooling flow model (B99; also see
David et al. 1994). Also evident is that the elliptical NGC 4636
$(\log_{10}L_{\rm x} = 41.53)$ has smaller $Z$ than NGC 1399 and NGC
4472 which have higher $L_{\rm x}$. Of the systems analyzed by B99 NGC
4636 is the only case which required the $\alpha$/Fe ratios to deviate
significantly from solar in order to achieve an acceptable fit, and it
had the largest contribution from a high energy component presumably
arising from discrete sources; i.e. the Fe abundance of NGC 4636
should be regarded as provisional.

The correlation of $Z$ with $L_{\rm x}$ (or with, e.g., $\dot{M}$)
still possesses considerable scatter, and thus there may be other
factors to consider. In particular, also in Figure \ref{fig.abungroup}
we plot $Z$ versus $R$, where $R$ is the radius of the circular
aperture used to extract the SIS0 spectrum for each object (see Table
\ref{tab.obs}). There is a clear correlation such that the
metallicities computed within the largest apertures tend to have the
smallest values.  For $R\sim 50$-100 kpc the correlation for the
groups has the least scatter; i.e. $Z$ falls from $\sim 1Z_{\sun}$ at
$R\sim 50$ kpc to $\sim 0.5Z_{\sun}$ at $R\sim 100$ kpc. At larger $R$
there is no evidence for further decline out to $R\sim 200$ kpc;
i.e. unlike the correlation with $L_{\rm x}$ it is not obvious from
these data that the $Z$-$R$ correlation joins smoothly with clusters.

Nevertheless, this trend of decreasing $Z$ with $R$ is consistent with
the presence of abundance gradients in these systems.  Since $R$ and
$L_{\rm x}$ are correlated to some extent because brighter sources
generally allow for larger extraction regions, we are unable at this
time to determine if the fundamental correlation of $Z$ is with
$L_{\rm x}$ or $R$ or whether they are equally important.  From
examination of Figure \ref{fig.abungroup} we reach the tentative
conclusion that $Z$ decreases due to increases in both $R$ and $L_{\rm
x}$, but for the groups in our sample abundance gradients are most
important for $R\sim 50$-100 kpc while the X-ray luminosity is the key
determinant for $\log_{10}L_{\rm x}\ga 42.5$. (For the ellipticals NGC
1399, NGC 4472, and NGC 4636 these effects cannot be distinguished.)

Renzini \shortcite{renzini} also shows that the gas fraction $(f_b)$
is approximately constant with temperature for clusters but it
plummets by several orders of magnitude for temperatures below $\sim
2$ keV. The determinations of $f_b$ are not extremely sensitive to
$Z_{\rm Fe}$ and thus our larger Fe abundances will not change the
trend in Renzini's Figure 4 by very much; i.e. the well-known
conclusion remains unchanged that the much smaller gas fractions of
groups imply that a significant portion of their gas be expelled
during their evolution in contrast to clusters.

The very sub-solar Fe abundances inferred from most previous studies
imply that $Z$ in groups is similar to or less than that of
clusters. Within the context of the ``standard chemical model'' which
assumes star formation properties similar to the Milky Way, these low
values of $Z$ require that groups accrete significant amounts of
primordial gas at late times after spending much of their evolution
expelling gas as demanded by their low $f_b$ (e.g. Renzini 1994,
1997). In this case the gas dynamical and metal enrichment history of
groups does not represent a smooth transition between ellipticals and
clusters neither of which are expected to accrete large amounts of
primordial gas at late times.  However, since our new results indicate
$Z\sim 0.7Z_{\sun}$ in groups the late accretion of primordial gas is
now not required, and thus the decrease of $Z$ from ellipticals to
groups and then to clusters can be understood in the context of the
``standard model'' as an increase in the proportion of primordial gas
retained as the mass of the system increases from ellipticals to
clusters.

The $\alpha$/Fe ratios obtained from our 2T and cooling flow models
are consistent with the solar values and those found for the brightest
ellipticals (BF; B99). This result taken with the nearly solar Fe
abundances indicates that, like clusters, the properties of the hot
gas in groups of galaxies are consistent with the Type Ia to Type II
supernova ratios and the IMF being similar to those of the Galaxy
(e.g. Renzini 1997; see also Fujita, Fukumoto, \& Okoshi 1997 and
Brighenti \& Mathews 1999b).

\section{Conclusions}
\label{conc}

We have analyzed the {\sl ASCA} spectra of 12 of the brightest groups
of galaxies previously studied by Davis et al. \shortcite{dmm} and
Mulchaey \& Zabludoff \shortcite{mz1132}. Because of the limitations
inherent in the {\sl ASCA} PSF we restricted our analysis to the
integrated spectra accumulated within a single circular aperture
corresponding to radii of $\sim 3\arcmin$-$5\arcmin$ ($\sim 50$-200
kpc) for the groups in our sample; these apertures typically enclose
$\sim 50$ per cent of the total X-ray emission.  Our principal
motivation for re-examining these data is to verify whether the very
sub-solar Fe abundances obtained from previous studies are in reality
an artifact of fitting isothermal models to the X-ray spectra as found
for bright elliptical galaxies by BF and B99.

Fitting isothermal models (1T) jointly to the {\sl ASCA} SIS and GIS
spectra gives results that are of poor or at best marginal quality for
most of the groups; i.e. $P\la 0.1$ where $P$ is the $\chi^2$ null
hypothesis probability. The primary cause of the poor fits is the
inability of the 1T models to fit the SIS data near 1 keV. Similar to
previous studies these 1T models give very sub-solar metallicities:
for the whole sample we obtain a mean and standard deviation $\langle
Z\rangle = 0.29\pm 0.12$~$Z_{\sun}$. Although allowing the
$\alpha$-process elements to vary separately from Fe often results in
$\alpha$/Fe ratios quite different from solar (especially for O and
Ne), the quality of the fits is not improved significantly in most
cases.  These properties of the 1T fits are very analogous to those of
the brightest ellipticals (see B99).

Two-temperature models (2T) provide significantly better fits than the
1T models for 11 out of the 12 groups in our sample. The superior
performance of the 2T models arises from their largely removing the
residuals in the SIS data near 1 keV present in the fits of the 1T
models. For groups where the temperature of the hotter component,
$T_{\rm h}$, is $<2$ keV the 2T models give $T_{\rm h}\approx 2T_{\rm
c}$ where $T_{\rm c}$ is the temperature of the colder component in
the hot gas. The ratio of emission measures of the two components is
$\rm EM_h/EM_c \sim 1$ for systems with the lowest average
temperatures and rises with increasing temperature until $\rm
EM_h/EM_c \sim 4$ for the hottest groups.

Although the column densities on the individual components are not as
precisely constrained as those for the brightest ellipticals (B99),
the impression is generally the same: $N_{\rm H}^{\rm c} > N_{\rm
H}^{\rm h}$ and $N_{\rm H}^{\rm h}$ is similar to the Galactic value.
We mention that the improvement of the 2T models over the 1T
models is not the result of the extra degree of freedom afforded by
separately varying $N_{\rm H}^{\rm c}$ and $N_{\rm H}^{\rm h}$, and
the temperatures and Fe abundances do not change substantially either. 

The metallicities of the 2T models are considerably larger than
obtained for the 1T models, $\langle Z\rangle = 0.75\pm
0.24$~$Z_{\sun}$; i.e. the mean metallicity is a factor of 2.6 larger
than the 1T value. Note that the 2T models have the $\alpha$/Fe ratios
fixed at their solar values. Even if the $\alpha$/Fe ratios are
allowed to be free parameters, the 2T models models do not indicate
Si/Fe ratios in excess of solar in contrast to previous studies using
isothermal models (e.g. Davis et al. 1999).

We obtain results for the multiphase cooling flow models that are
entirely analogous to those of the 2T models including the large
metallicities, $\langle Z\rangle = 0.65\pm 0.17$~$Z_{\sun}$. Typically
the mass deposition rates $(\dot{M})$ are 5-10 $M_{\sun}$ yr$^{-1}$
placing the groups midway between ellipticals and clusters
(e.g. Fabian 1994) as expected. Hence, the value of $\dot{M}$ for the
cooling flow models, or equivalently the emission measure of the
cooler component for the 2T models, are viable means of distinguishing
groups from ellipticals and clusters.

In section \ref{prev} we have presented an extensive comparison of our
results to previous studies. We conclude that the very sub-solar Fe
abundances and Si/Fe enhancements in groups of galaxies found by most
previous studies (e.g. Fukazawa et al. 1996; Davis et al. 1999) are an
artifact of their fitting isothermal models to intrinsically
multiphase {\sl ASCA} spectra. (We also critically evaluate some
studies of spatially resolved spectral analysis.) These results are
virtually identical to those found for the brightest elliptical
galaxies by BF and B99. Owing to the importance of these results for
interpreting X-ray spectra, in Appendix \ref{bias} we use simulated
{\sl ASCA} observations to explore in some detail the ``Fe bias'' and
``Si bias'' associated with the spectral fitting of ellipticals,
groups, and clusters of galaxies.

We have discussed how the properties of the hot gas in groups of
galaxies within the central regions ($r\sim 100$ kpc) are revised in
light of the 2T and cooling flow models, and how the properties are
consistently intermediate with those of ellipticals and clusters
(sections \ref{prop} and \ref{abun}). Also considered are physical
scenarios for the 2T models and how they can be distinguished from
cooling flows with future X-ray missions (section
\ref{2tvscf}). Finally, we examined the implications of the nearly
solar Fe abundances and solar $\alpha$/Fe ratios for the groups and
how they are consistent with popular chemical models.

Our results for the brightest ellipticals (BF; B99) and the brightest
groups (this paper) demonstrate the similarity of the properties of
the hot gas in the centers of these systems to cooling flow (i.e. cD)
clusters (e.g. Fabian 1994). The different Fe abundances measured for
these systems may reflect the sizes of the apertures used for spectral
analysis; i.e. the physical aperture size typically used when
measuring $Z_{\rm Fe}$ increases from approximately 30 kpc for
ellipticals to 100 kpc for groups up to at least 200-300 kpc for
clusters.  However, presently the role of abundance gradients cannot
be disentangled from that of the X-ray luminosity (or $\dot{M}$) of
the system (see section \ref{abun}).  Although ellipticals and groups
are believed to expel much of their hot gas during their evolution
while clusters are ``closed boxes'' and retain all of their gas, these
differences may not much affect the gas properties in their centers,
and thus the strong similarity of the gas properties in the cores of
these systems is not unreasonable.

The role of abundance gradients in ellipticals, groups, and clusters
should be clarified significantly with data from the upcoming {\sl
Chandra} and {\sl XMM} satellites which combine high spatial and
spectral resolution unlike previous missions.

\section*{Acknowledgments}

We thank W. Mathews for interesting discussions and for comments on
the manuscript. This research has made use of (1) ASCA data obtained
from the High Energy Astrophysics Science Archive Research Center
(HEASARC), provided by NASA's Goddard Space Flight Center, and (2) the
NASA/IPAC Extragalactic Database (NED) which is operated by the Jet
Propulsion Laboratory, California Institute of Technology, under
contract with the National Aeronautics and Space Administration.  The
{\sc XSPEC} implementation of the multiphase cooling flow model was
kindly provided by R. Johnstone to whom we express our
gratitude. Support for this work was provided by NASA through Chandra
Fellowship grant PF8-10001 awarded by the Chandra Science Center, which
is operated by the Smithsonian Astrophysical Observatory for NASA
under contract NAS8-39073.

\appendix

\section{Biases associated with X-ray spectral fitting}  
\label{bias}

\begin{table*}
\caption{Models}
\label{tab.models}
\begin{tabular}{ccccccccc}
Model & Type &  $N_{\rm H}^{\rm c}$ & $N_{\rm H}^{\rm h}$ & $T_{\rm
     c}$ & $T_{\rm h}$ & $Z$ & $\rm EM_c$ or $\dot{M}$ & $\rm EM_h$\\   
      & & ($10^{21}$ cm$^{-2}$) & ($10^{21}$ cm$^{-2}$)  & (keV) &
     (keV) & $(Z_{\sun})$ & 
     (see notes) & (see notes)\\   
1 & 2T    & 2.5 & 0.5 & 0.75 & 1.5 & 1.0 & 2.0 & 2.0\\
2 & CF+1T & 2.5 & 0.5 & 1.5 & tied & 1.0 & 1.5 & 2.0\\
3 & CF+1T & 2.5 & 0.5 & 2.5 & tied & 1.0 & 10.0 & 1.5\\
4 & CF+1T & 6.8 & 2.2 & 6.8 & tied & 0.34 & 653 & 22.4\\

\end{tabular}
\medskip

\raggedright

The units of emission measure ($\rm EM$) are defined as in the notes
to Table \ref{tab.fits}, and the mass deposition rates $(\dot{M})$ are
listed in units of $M_{\sun}$ yr$^{-1}$. See text for further
explanation of these models.

\end{table*}

\begin{table*}
\caption{1T fits to simulated data}
\label{tab.1tvar}
\begin{tabular}{ccccccccc}
Model & $N_{\rm H}$ & $T$ & $Z_{\rm Fe}$ & $Z_{\rm O}$ & $Z_{\rm Ne}$
      & $Z_{\rm Mg}$  & $Z_{\rm Si}$ & $Z_{\rm S}$\\    
      & ($10^{21}$ cm$^{-2}$) & (keV) & $(Z_{\sun})$ & $(Z_{\sun})$ &
      $(Z_{\sun})$ & $(Z_{\sun})$ & $(Z_{\sun})$ & $(Z_{\sun})$\\ 
1 & 0.74 & 1.03 & 0.26 & 0.00 & 0.00 & 0.15 & 0.30 & 0.42\\
2 & 0.35 & 1.16 & 0.31 & 0.00 & 0.00 & 0.27 & 0.32 & 0.45\\
3 & 0.30 & 2.20 & 0.66 & 0.00 & 1.39 & 0.67 & 0.71 & 0.79\\
4 & 3.47 & 5.00 & 0.33 & 2.34 & 0.94 & 0.44 & 0.64 & 0.55\\

\end{tabular}
\medskip

\raggedright

Best-fitting 1T model parameters resulting from fits to simulated {\sl
ASCA} data of the models in Table \ref{tab.models}. All elements not
listed have abundances tied to Fe in their solar ratios. See text for
discussion of the {\sl ASCA} simulations.

\end{table*}

\begin{figure*}
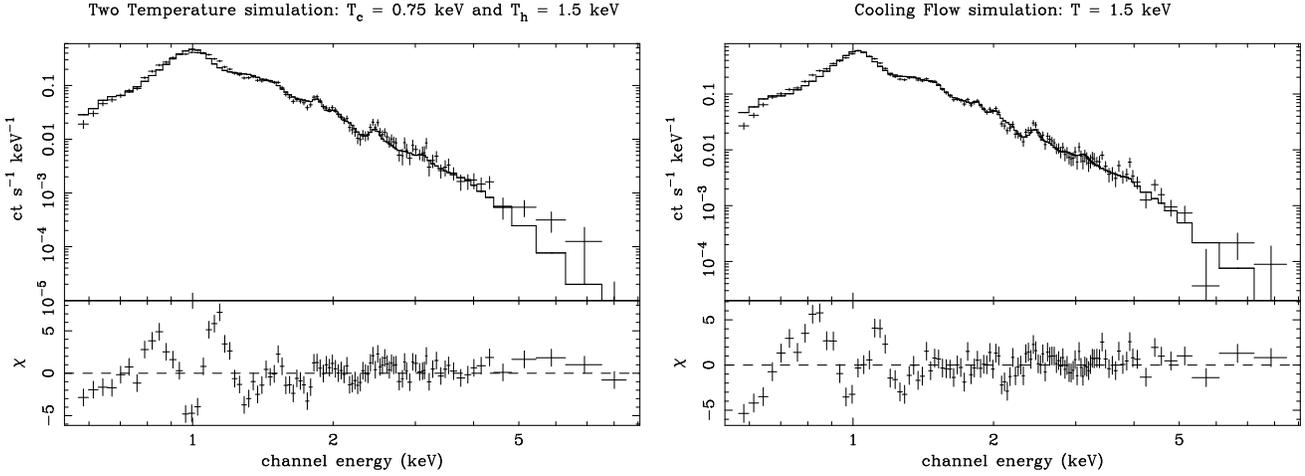

\parbox{0.49\textwidth}{
\centerline{\psfig{figure=2tsim_n1399.ps,angle=-90,height=0.26\textheight}}
}
\parbox{0.49\textwidth}{
\centerline{\psfig{figure=cfsim_n1399.ps,angle=-90,height=0.26\textheight}}
}
\caption{\label{fig.sim.n1399} Best-fitting 1T models and residuals
for the simulated {\sl ASCA} data of models 1 (left) and 2 (right)
listed in Table \ref{tab.models}. These 1T fits have the $\alpha$/Fe
ratios fixed at the solar values to highlight the residuals near the
regions of the strong emission lines of the $\alpha$-process elements.}

\end{figure*}

\begin{figure*}
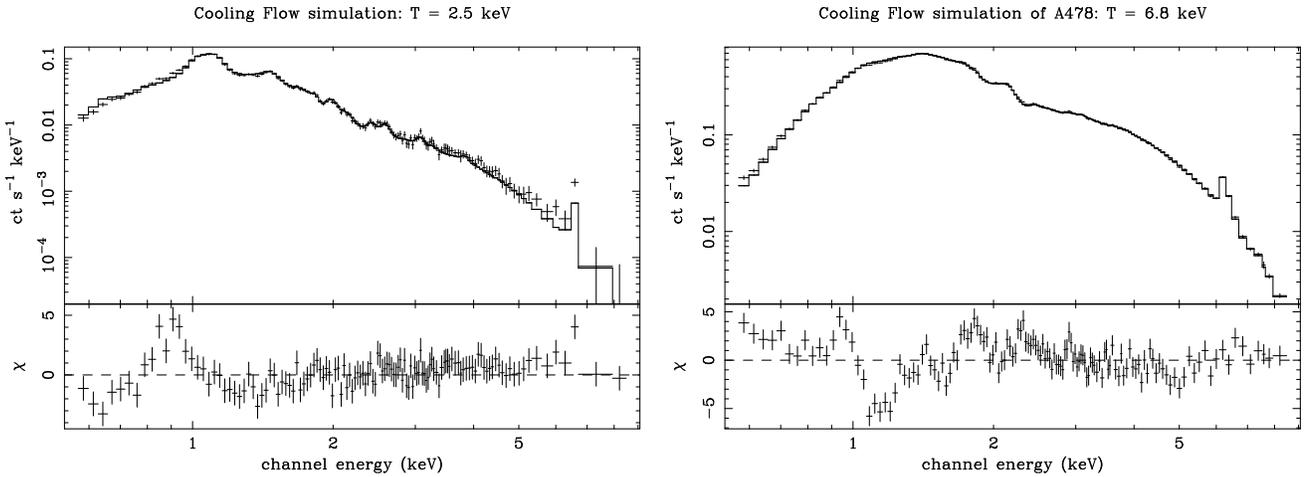

\parbox{0.49\textwidth}{
\centerline{\psfig{figure=cfsim_mkw9.ps,angle=-90,height=0.26\textheight}}
}
\parbox{0.49\textwidth}{
\centerline{\psfig{figure=cfsim_a478.ps,angle=-90,height=0.26\textheight}}
}
\caption{\label{fig.sim.hight} As Figure \ref{fig.sim.n1399} except we
display results for the simulated {\sl ASCA} data of models 3 (left)
and 4 (right) of Table \ref{tab.models}.}

\end{figure*}

The X-ray studies of the hot gas in elliptical galaxies NGC 720 by
Buote \& Canizares \shortcite{bc94} and NGC 4636 by Trinchieri et
al. \shortcite{trin} were the first to show that the {\sl ROSAT} PSPC
spectra of ellipticals could not distinguish between isothermal models
with very sub-solar Fe abundances and two-temperature models with
solar Fe abundances. Using simulated PSPC data Buote \& Canizares
\shortcite{bc94} further demonstrated that if the galaxy spectrum is
intrinsically characterized by two temperatures in the hot gas with
solar Fe abundances but is instead fitted with an isothermal model
then sub-solar Fe abundances necessarily result; i.e. the isothermal
model gives a strongly biased measurement of the Fe
abundance\footnote{We note that Bauer \& Bregman \shortcite{bb}
discuss difficulties with determining metallicities from PSPC
observations but not with respect to a fitting bias resulting from
assuming an isothermal model for elliptical galaxies. Kim et
al. \shortcite{kim96} show that the metallicity inferred from spectral
fitting of the {\sl ASCA} and {\sl ROSAT} data of the faint S0 galaxy
NGC 4382 is essentially zero for isothermal models, but it cannot be
constrained (and is thus consistent with solar) by a two-component
model with $T_1\sim 0.3$ keV for the hot gas and $T_2\sim 5$ keV for a
bremsstrahlung component. Although this shows the degeneracies in
metallicity depending on the spectral model of a faint (low $L_{\rm
x}/L_{\rm B}$) elliptical, this does not demonstrate a bias resulting
from fitting a spectrum with two temperatures in the hot gas
(i.e. with $T\sim 0.5$-2 keV) apart from any bremsstrahlung
component.}.

The strong dependence of the Fe abundance on spectral model has now
been demonstrated using {\sl ASCA} data for elliptical galaxies by BF
and B99 and for groups of galaxies in this paper. These papers also
show that multiphase models are clearly favored over isothermal models
for these systems.  As a result, investigators hoping to obtain
average Fe abundances for these objects by fitting isothermal models
in actuality always infer strongly biased (i.e. very sub-solar) Fe
abundances for ellipticals (e.g. Matsumoto et al. 1997) and for groups
(e.g Davis et al. 1999).  Since this ``Fe bias'' is crucial for
interpreting the X-ray spectra of these systems, in this Appendix we
attempt to elucidate this bias by using simulations of simple
multiphase models that are consistent with the current {\sl ASCA}
observations of ellipticals, groups, and clusters.

We investigate the dependence of the Fe bias on temperature by
considering objects spanning the temperatures of the brightest
ellipticals, groups, and clusters of galaxies. In addition, we use
{\sl ASCA} simulations of these systems to show the dependence of the
$\alpha$/Fe ratios on the spectral model. Thus, this Appendix updates
and extends the brief studies of the Fe bias that appear in section
2.2 of Buote \& Canizares \shortcite{bc94} and in section 3.2.2 of BF.

The key properties of the four reference models we use for this study
are listed in Table \ref{tab.models}. Each of these models corresponds
to regions of positive temperature gradient (e.g. see section 5 of
B99).  Models 1 and 2 are based on the two-temperature and cooling
flow models of the elliptical galaxy NGC 1399 within $r\sim 30$ kpc
(B99) which are also quite similar to the corresponding models of
several of the lower-temperature groups in this paper (see sections
\ref{2t} and \ref{cf}). To explore the high-temperature groups we base
model 3 on the cooling flow models of MKW 9 and NGC 4104 (section
\ref{cf}).

Finally, we obtained the {\sl ASCA} data of A478 from the HEASARC
archive and have analyzed the spectra within a $3\arcmin$
radius\footnote{This radius corresponds to 342 kpc assuming a redshift
of 0.0882, $H_0=70$ km s$^{-1}$ Mpc$^{-1}$, and $\Omega_0=0.3$.} in a
similar manner to the ellipticals and groups (see section \ref{obs}):
the result is listed as model 4 and is consistent with previous
studies (e.g. Johnstone et al. 1992). A478 was selected because (1) it
is one of the brightest clusters having evidence for substantial
excess absorption and (2) with respect to other cooling flow clusters
a comparatively large fraction of the emission measure arises from the
cooling flow component; i.e. A478 should be one of the most favorable
clusters for examining systematics when fitting an isothermal model.

We simulated {\sl ASCA} observations of the models in Table
\ref{tab.models} using {\sc XSPEC}. Because of its better energy
resolution and sensitivity near 1 keV we focus on the SIS data. For
consistency of presentation we use the same PI grouping for each
simulated SIS spectrum (i.e. that of the SIS data of NGC 1399 -- see
B99).  Our choices of exposure times for the simulations are motivated
by a desire to insure (1) sufficiently high S/N for each case so that
noise does not noticeably affect the fitted parameter values, (2) that
the sizes of the residuals (in terms of $\chi^2$) are similar to the
best available {\sl ASCA} data, and (3) that the simulated data for
each model have similar S/N. Although background is included in the
simulations, the S/N criteria stated above insure that its effect is
only noticeable at the highest energies which are not of interest to
our study.

\subsection{The Fe bias}
\label{fe}

In Figure \ref{fig.sim.n1399} we plot the simulated {\sl ASCA} SIS
data of models 1 and 2 from Table \ref{tab.models} and also show the
best-fitting isothermal model and its residuals. These 1T models with
the $\alpha$/Fe ratios fixed at solar give best-fitting values of
$Z=0.27Z_{\sun}$ for the simulated 2T model and $Z=0.32Z_{\sun}$ for
the simulated CF+1T model. (These metallicities do not differ
significantly from the Fe abundances obtained when the $\alpha$/Fe
ratios are allowed to vary in the fits -- see Table \ref{tab.1tvar}.)
Notice that both the pattern of residuals and the metallicities
derived from the 1T fits are very similar for both models 1 and 2 and
thus the Fe bias is essentially the same for isothermal fits to
two-temperature and cooling flow spectra.

This Fe bias is insensitive to the presence of excess absorption on
the colder component. If instead we consider a model identical to
model 1 but with $N_{\rm H}^{\rm c}=N_{\rm H}^{\rm h}= 5\times
10^{20}$ cm$^{-2}$, then the metallicity obtained from fitting the 1T
model to the simulated {\sl ASCA} data has a best-fitting value
$Z=0.25Z_{\sun}$ which is consistent with the value obtained from the
simulated model 1 within the $1\sigma$ errors. Below 1 keV the pattern
of residuals for this model differs from model 1 such that the
residual peak below 1 keV is slightly less pronounced than in Figure
\ref{fig.sim.n1399}.

The pattern of residuals of model 3 (Figure \ref{fig.sim.hight}) is
less complex than for the lower temperature models 1 and 2.  The
best-fitting metallicity using the 1T model ($\alpha$/Fe fixed at
solar) is $0.61Z_{\sun}$ which represents only a 40\% under-estimate
of the true value in contrast to the $\sim 70\%$ under-estimate
observed for models 1 and 2. The Fe bias decreases more dramatically
as we move to still higher temperatures with model 4 (Figure
\ref{fig.sim.hight}). For this model we obtain a best-fitting value of
$Z=0.30Z_{\sun}$ using the 1T model ($\alpha$/Fe fixed at solar) which
is only a $10\%$ under-estimate of the true value. If we allow the
$\alpha$/Fe ratios to vary (Table \ref{tab.1tvar}) the best-fitting Fe
abundance rises to $0.33Z_{\sun}$; i.e. the Fe bias for hot clusters
is insignificant.

A physical explanation of the Fe bias is as follows. For models 1 and
2 which have average temperatures $\sim 1$ keV, the colder temperature
components preferentially excite emission lines in the Fe L complex
below 1 keV (from ions Fe {\romannumeral 17} - {\romannumeral 21})
while the hotter temperature components preferentially excite the Fe L
lines from $\sim 1$-1.4 keV (Fe {\romannumeral 21} - {\romannumeral
24}). Since these components contribute approximately equally to the
emission measure, the net result is that the spectral shape in the
highly temperature-sensitive Fe L energy region is flattened with
respect to a 1T model. That is, the spectrum of a 1T model which has
the average temperature of the two components of models 1 or 2 (and
has the same Fe abundance) will preferentially excite the Fe L lines
right at 1 keV at the expense of the other Fe L lines near $\sim 0.7$
keV and $\sim 1.3$ keV and thus will be more peaked at 1 keV than
either model 1 or 2.  To partially compensate for this, i.e. to
flatten the spectral shape near 1 keV, the 1T model decreases its Fe
abundance so that the contribution of the flatter bremsstrahlung
continuum is increased in the model. However, the distinctive residual
pattern in Figure \ref{fig.sim.n1399} demonstrates that the Fe L lines
are sufficiently temperature sensitive (and the SIS has sufficient
energy resolution) to thwart this attempt to successfully fit the 1T
model to the simulated two-temperature and cooling flow spectra.

The strong temperature-sensitive Fe L emission lines are the key to
this Fe bias. As one moves to higher temperatures (models 3 and 4) the
Fe L lines become weaker and the bias decreases rapidly for average
temperatures above 2 keV. Moreover, at higher temperatures the
strong 6.5 keV Fe K emission line complex becomes increasingly more
important and dominates the determination of the Fe abundance.

A final note: since it is the variation of the spectral shape of the
Fe L complex with temperature which gives rise to the Fe bias, one
does not obtain a biased measurement of the Fe abundance using a 1T
model if the spectrum consists of multiple spectral components which
have the same temperature but different Fe abundances. In this case
the 1T model yields a value for the Fe abundance that is a weighted
average over the various components.

\subsection{The Si and S bias}
\label{si}

Prominent residuals in the 1T fits ($\alpha$/Fe fixed at solar) are
located near the strong emission lines of O, Ne, Mg, Si, and S
(Figures \ref{fig.sim.n1399} and \ref{fig.sim.hight}). If the
abundances of these elements are allowed to vary separately from Fe we
obtain the results listed in Table \ref{tab.1tvar}. Although in most
instances the O, Ne, and (to a lesser extent) the Mg abundances
differ significantly from Fe, their new values do not correspond to a
substantial reduction of the residuals over $\sim 0.5$-1.5 keV.

The 1T fits for models 1 and 2 produce zero O and Ne abundances and
Mg/Fe ratios less than solar. These results are very sensitive to the
presence of excess absorption on the colder temperature component.  If
instead we consider a model identical to model 1 but with $N_{\rm
H}^{\rm c}=N_{\rm H}^{\rm h}= 5\times 10^{20}$ cm$^{-2}$, then we
obtain $Z_{\rm O}=0.55Z_{\sun}$, $Z_{\rm Ne}=1.1Z_{\sun}$, and $Z_{\rm
Mg}=0.23Z_{\sun}$ with $Z_{\rm Fe}=0.26Z_{\sun}$.

For models 3 and 4 the 1T fits yield O, Ne, and Mg abundances very
different from models 1 and 2 and from each other. Although the 1T fit
for model 3 gives zero O abundance its large Ne abundance is
qualitatively different from models 1 and 2. The ratios of O/Fe,
Ne/Fe, and Mg/Fe obtained for model 4 are all very much in excess of
solar. Hence, the O, Ne, and Mg abundances are very sensitive to the
temperature model and to the absorption model, and thus the systematic
errors introduced when using a 1T model to measure these abundances
are more complex in origin than the Fe bias discussed above.

In contrast, the K$\alpha$ emission lines of Si and S are at
sufficiently high energies so that they are not affected by the
absorption model. The residuals near the Si ($\sim 1.8$-2.0 keV) and S
($\sim 2.4$-2.7 keV) lines are readily apparent for all the models in
Figures \ref{fig.sim.n1399} and \ref{fig.sim.hight}, especially for
model 4. The 1T fits with variable $\alpha$/Fe ratios listed in Table
\ref{tab.1tvar} for models 1-3 indicate modest ratios of $\la 1.1$ for
$Z_{\rm Si}/Z_{\rm Fe}$ and more substantial ratios of $\sim 1.2$-1.5
solar for $Z_{\rm S}/Z_{\rm Fe}$. Model 4 has the largest ratios with
$Z_{\rm Si}/Z_{\rm Fe}\sim 2$ and $Z_{\rm S}/Z_{\rm Fe}\sim
1.7$. Thus, the inferred values of Si/Fe tend to increase with
temperature while the S/Fe ratio is relatively constant.
  
These excess Si/Fe and S/Fe ratios inferred by 1T models arise from
two effects: (1) the colder temperature components present in the
two-component models  produce stronger Si and S emission, and (2)
the 1T model under-estimates the continuum at energies above $\sim 2$
keV. Issue \#1 is more important for the higher temperature systems
like clusters (model 4) while issue \#2 dominates for the lower
temperature systems like elliptical galaxies (models 1-2).

Therefore, attempts to infer the Si/Fe and S/Fe ratios using
isothermal models on spectra that are intrinsically composed of
multiple temperature components will necessarily yield values for
these ratios that are biased in excess of the true values, the effect
of which is most important for higher temperature clusters of
galaxies. This bias may have played a role in the excess Si/Fe ratios
obtained by Mushotzky et al. \shortcite{mush96} and may in part
explain the trend discovered by Fukazawa et al. \shortcite{fuk98} that
the Si/Fe ratio rises with temperature for clusters galaxies.
Although Fukazawa et al. attempt to eliminate any problems associated
with multiphase gas by excluding the central regions of their
clusters, it is quite possible that their arbitrary method of defining
the affected regions does not fully remove the effects of multiphase
gas and the resulting Si bias we have described.

\end{document}